\documentclass[aps,prc,twocolumn,superscriptaddress,showpacs,showkeys,nofootinbib,floatfix]{revtex4}
\usepackage{amssymb,latexsym, amsmath, graphicx}  
\usepackage{longtable}

\newcommand{\cm}[2][\,]{\ensuremath{#2^{#1\ast}}}
\newcommand{\im}[1]{\Im\left(#1\right)}
\newcommand{\re}[1]{\Re\left(#1\right)}

\newcommand{\cgln}{\frac{2W|\cm{\mathbf{k}}_{\pi}|}{W^2-m_p^2}}
\newcommand{\eq}{Eq.}
\newcommand{\peq}{Eqs.}
\newcommand{\fig}{Fig.}
\newcommand{\pfig}{Figs.}
\newcommand{\tab}{Tbl.}
\newcommand{\ptab}{Tbls.}
\newcommand{\sref}{Ref.}
\newcommand{\pref}{Refs.}

\newcommand{\sect}{Sec.}

\newcommand{\qfac}{\sqrt{\frac{2 Q^2}{|\mathbf{q}^{\ast}|^2}}}

\newcommand{\umasss}{(GeV/c$^2$)$^2$}
\newcommand{\umom}{GeV/c}

\newcommand{\uenergy}{GeV}

\newcommand{\ufourmomts}{(GeV/c$^2$)$^2$}

\begin{document}


\title{Neutral pion electroproduction in the resonance \\ region
at high $Q^2$ }


\author{A.N.~Villano}
\email[Electronic addresses:]{villaa@jlab.org,villaa@umich.edu}
\altaffiliation[Present address: ]{
University of Michigan, 
Ann Arbor MI 48109}
\affiliation{Rensselaer Polytechnic Institute, Troy, New York 12180}

\author{P.~Stoler}
\affiliation{Rensselaer Polytechnic Institute, Troy, New York 12180}
\author{P.E.~Bosted}
\affiliation{Thomas Jefferson National Accelerator Facility, Newport News, Virginia 23606}
\author{S.H.~Connell}
\affiliation{University of the Johannesburg, Johannesburg, South Africa}
\author{M.M.~Dalton}
\affiliation{University of the Witwatersrand, Johannesburg, South Africa}
\author{M.K.~Jones}
\affiliation{Thomas Jefferson National Accelerator Facility, Newport News, Virginia 23606}
\author{V.~Kubarovsky}
\affiliation{Rensselaer Polytechnic Institute, Troy, New York 12180}
\author{G.S.~Adams}
\affiliation{Rensselaer Polytechnic Institute, Troy, New York 12180}
\author{A.~Ahmidouch}
\affiliation{North Carolina A \& T State University, Greensboro, North Carolina 27411}
\author{J.~Arrington}
\affiliation{Physics Division, Argonne National Laboratory, Argonne, Illinois 60439}
\author{R.~Asaturyan}
\thanks{deceased}
\affiliation{Yerevan Physics Institute, Yerevan, Armenia}
\author{O.K.~Baker}
\affiliation{Hampton University, Hampton, Virginia 23668}
\affiliation{Thomas Jefferson National Accelerator Facility, Newport News, Virginia 23606}
\author{H.~Breuer}
\affiliation{University of Maryland, College Park, Maryland 20742}
\author{M.E.~Christy}
\affiliation{Hampton University, Hampton, Virginia 23668}
\author{S.~Danagoulian}
\affiliation{North Carolina A \& T State University, Greensboro, North Carolina 27411}
\author{D.~Day}
\affiliation{University of Virginia, Charlottesville, Virginia 22901}
\author{J.A.~Dunne}
\affiliation{Mississippi State University, Mississippi State, Mississippi 39762}
\author{D.~Dutta}
\affiliation{Mississippi State University, Mississippi State, Mississippi 39762}
\affiliation{Triangle Universities Nuclear Laboratory and Duke University, Durham, North Carolina 27708}
\author{R.~Ent}
\affiliation{Thomas Jefferson National Accelerator Facility, Newport News, Virginia 23606}
\author{H.C.~Fenker}
\affiliation{Thomas Jefferson National Accelerator Facility, Newport News, Virginia 23606}
\author{V.V.~Frolov}
\affiliation{LIGO Livingston Observatory, Livingston, LA 70754}
\author{L.~Gan}
\affiliation{University of North Carolina Wilmington, Wilmington, North Carolina 28403}
\author{D.~Gaskell}
\affiliation{Thomas Jefferson National Accelerator Facility, Newport News, Virginia 23606}
\author{W.~Hinton}
\affiliation{Hampton University, Hampton, Virginia 23668}
\affiliation{Thomas Jefferson National Accelerator Facility, Newport News, Virginia 23606}
\author{R.J.~Holt}
\affiliation{Physics Division, Argonne National Laboratory, Argonne, Illinois 60439}
\author{T.~Horn}
\affiliation{University of Maryland, College Park, Maryland 20742}
\author{G.M.~Huber}
\affiliation{University of Regina, Regina, Saskatchewan, Canada, S4S 0A2}
\author{K.~Joo}
\affiliation{University of Connecticut, Storrs, Connecticut 06269}
\author{N.~Kalantarians}
\affiliation{University of Houston, Houston, TX 77204}
\author{C.E.~Keppel}
\affiliation{Hampton University, Hampton, Virginia 23668}
\affiliation{Thomas Jefferson National Accelerator Facility, Newport News, Virginia 23606}
\author{Y.~Li}
\affiliation{University of Houston, Houston, TX 77204}
\author{A.~Lung}
\affiliation{Thomas Jefferson National Accelerator Facility, Newport News, Virginia 23606}
\author{D.~Mack}
\affiliation{Thomas Jefferson National Accelerator Facility, Newport News, Virginia 23606}
\author{S.~Malace}
\affiliation{Bucharest University, Bucharest, Romania}
\author{P.~Markowitz}
\affiliation{Florida International University, University Park, Florida 33199}
\author{D.G.~Meekins}
\affiliation{Thomas Jefferson National Accelerator Facility, Newport News, Virginia 23606}
\author{H.~Mkrtchyan}
\affiliation{Yerevan Physics Institute, Yerevan, Armenia}
\author{J.~Napolitano}
\affiliation{Rensselaer Polytechnic Institute, Troy, New York 12180}
\author{G.~Niculescu}
\affiliation{University of Virginia, Charlottesville, Virginia 22901}
\author{I.~Niculescu}
\affiliation{James Madison University, Harrisonburg, Virginia 22807}
\author{D.H.~Potterveld}
\affiliation{Physics Division, Argonne National Laboratory, Argonne, Illinois 60439}
\author{Paul~E.~Reimer}
\affiliation{Physics Division, Argonne National Laboratory, Argonne, Illinois 60439}
\author{J.~Reinhold}
\affiliation{Florida International University, University Park, Florida 33199}
\author{J.~Roche}
\affiliation{Thomas Jefferson National Accelerator Facility, Newport News, Virginia 23606}
\author{S.E.~Rock}
\affiliation{University of Massachusetts Amherst, Amherst, Massachusetts 01003}
\author{G.R.~Smith}
\affiliation{Thomas Jefferson National Accelerator Facility, Newport News, Virginia 23606}
\author{S.~Stepanyan}
\affiliation{Yerevan Physics Institute, Yerevan, Armenia}
\author{V.~Tadevosyan}
\affiliation{Yerevan Physics Institute, Yerevan, Armenia}
\author{V.~Tvaskis}
\affiliation{Dept. of Physics, VU University, NL-1081 HV Amsterdam, The Netherlands}
\author{M.~Ungaro}
\affiliation{University of Connecticut, Storrs, Connecticut 06269}
\affiliation{Thomas Jefferson National Accelerator Facility, Newport News, Virginia 23606}
\author{A.~Uzzle}
\affiliation{Hampton University, Hampton, Virginia 23668}
\author{S.~Vidakovic}
\affiliation{University of Regina, Regina, Saskatchewan, Canada, S4S 0A2}
\author{F.R.~Wesselmann}
\affiliation{University of Virginia, Charlottesville, Virginia 22901}
\author{B.~Wojtsekhowski}
\affiliation{Thomas Jefferson National Accelerator Facility, Newport News, Virginia 23606}
\author{S.A.~Wood}
\affiliation{Thomas Jefferson National Accelerator Facility, Newport News, Virginia 23606}
\author{L.~Yuan}
\affiliation{Hampton University, Hampton, Virginia 23668}
\author{X.~Zheng}
\affiliation{Physics Division, Argonne National Laboratory, Argonne, Illinois 60439}
\author{H.~Zhu}
\affiliation{University of Virginia, Charlottesville, Virginia 22901}


\date{\today}

\begin{abstract}
The process $ep \rightarrow ep\pi^0$ has been measured at $Q^2$ = 6.4 and 7.7~\ufourmomts \ in Jefferson Lab's Hall C.  Unpolarized differential cross sections
are reported in the virtual photon-proton center of mass frame considering the process $\gamma^{\ast}p \rightarrow p\pi^0$.  Various details
relating to the background subtractions, radiative corrections and systematic errors are discussed.  The usefulness of the data with
regard to the measurement of the electromagnetic properties of the well known $\Delta(1232)$ resonance is covered in detail.  Specifically
considered are the electromagnetic and scalar-magnetic ratios $R_{EM}$ and $R_{SM}$ along with the magnetic transition form factor $G_M^{\ast}$.  It
is found that the rapid fall off of the $\Delta(1232)$ contribution continues into this region of momentum transfer and that other resonances
may be making important contributions in this region.
\end{abstract}


\maketitle

\section{\label{S:physics}Physical Motivation}
Electromagnetic elastic and transition form factors have historically proved  essential 
in furthering the understanding of baryon structure  and the concomitant degrees of freedom
necessary to describe it. The spectra of baryon transition resonances led directly to the
quark model, and the basic measurable static and dynamic  properties of many excited 
baryon states  were successfully described by the constituent quark model (CQM).
Properties of charge and current distributions 
such as the charge radius were obtained from elastic electron scattering as a function of
the 4-momentum transfer $q^2$.  

By far the most studied of the resonances has been the $\Delta(1232)$, which has both  spin and 
isospin  quantum numbers of 3/2. It is the lowest lying excitation and  it decays almost exclusively into the
simple $N-\pi$\  final state with  a  $p$-wave.  It is 
relatively isolated from other resonances and is very strongly excited, almost completely 
saturating the unitary circle in an Argand plot. Since its spin is 3/2 it can be electromagnetically
excited via three electromagnetic  multipoles - $M1$, $E2$ and $S1$, which denote magnetic dipole,
electric quadrupole and scalar dipole, respectively. 

For real photons ($Q^2$=0) the $\Delta(1232)$ resonance (hereafter simply referred to as ``the $\Delta$'') is  nearly a pure $M1$ excitation.
Early on this was explained  in the framework of the  SU(6) CQM as a magnetic spin-flip 
excitation of one of the nucleon's  quarks,  which move in a spherically symmetric oscillator type 
potential~\cite{Becchi:1965vo}.  However, it is found  that the $\Delta$ excitation 
also has small, but non-zero, components of $E2$ and $S1$ amplitudes.  
Near $Q^2$=0 it is found that the ratio $R_{EM} \equiv E2/M1 \sim$ -0.02 to -0.03 . This  non-zero  
$R_{EM} $  implies that the $N\to\Delta$ transition
has an  electric quadrupole moment and therefore the $\Delta$ is slightly deformed from 
sphericity.  The splitting of the $\Delta$ mass from the nucleon has been 
interpreted~\cite{Isgur:1978xj, Isgur:1981yz} as arising from  a color hyperfine interaction, which also 
induces the small electric quadrupole moment.
The existence of this small distortion of shape has been alternatively described~\cite{Sato:1996gk,Sato:2000jf} 
as a non-spherical pion cloud, which is part of the sea quarks, 
surrounding the spherical quark core.

As $Q^2$ increases one begins to penetrate this cloud and access the core. The  small
wavelength virtual photons begin to resolve current quarks. The description of the process must evolve
with with $Q^2$ as well. At the asymptotic limit, $Q^2\to\infty$,  it
is widely accepted that the pQCD approach should explain all exclusive reactions in which the entire process involves only the  minimum Fock state configuration of quarks, which  exchange  the minimum number of gluons.
For baryon elastic and transition form factors  this implies three valence quarks exchanging two gluons, with helicity  conservation at each vertex.  The result is the so-called pQCD
 {\em  constituent scaling}, which for baryons means the leading  form
factors should scale as $1/Q^4$.  In addition to constituent scaling, 
the pQCD process requires helicity conservation for the overall process. 
 
The question of how to describe exclusive reactions at $Q^2$ between zero and infinity is 
one of the major fields of study in nuclear physics today, and will be continue to be so in the foreseeable
future.  The present range of $Q^2$ over which baryon  form factors can  
be studied  in detail (aside from the elastic proton magnetic form factor $G_{Mp}$) 
is approximately from $0$ to around 8~\ufourmomts, over which the wavelength of the probe
varies from about 1~fm to less than 0.05~fm. Over such a large range of probe resolution 
 it is not clear which models of description are most appropriate, and their ranges of 
 relevance must also evolve.

The present analysis is concerned with the upper range of the available  momentum transfers. 
There are several approaches which have been applied to the study of the exclusive reactions and 
baryon form factors in this kinematic range:  pQCD; generalized parton distributions (GPD); light cone-sum rules (LCSR);  lattice QCD (LQCD); and  relativistic versions of the CQM.
A review of the physics of resonances  at high $Q^2$ can be found in \sref~\cite{Burkert:09ts}, which also includes pertinent references.
The important signatures relating to the onset of pQCD are the constituent  scaling 
rules and helicity conservation. 
The scaling  rules predict that the leading order $N\to\Delta$ 
transition form factor  $G_M^{\ast}$, which is directly related to
the dominant $M_{1+}$ multipole,  scales as $1/Q^4$. 
Helicity conservation implies $R_{EM}=+1$. A further consequence of pQCD is that $R_{SM}$ be a constant.
It would be very significant if $G_M^{\ast}$, $R_{EM}$, and $R_{SM}$
begin to approach these behaviors in the range $2.5 \le Q^2 \le 10.0$~\ufourmomts.
At intermediate values of $Q^2$ estimates have been
made in terms of GPDs~\cite{Stoler:1993yk}, LCSRs~\cite{Stoler:1993yk}
large $N_C$ and chiral limits~\cite{Pascalutsa:2006ne,Pascalutsa:2006up}, and LQCD~\cite{Alexandrou:2004xn}.

Earlier analysis of inclusive electron scattering data at SLAC~\cite{Stoler:1993yk,Stuart:1996zs}
indicated that the $p\to\Delta$ form factor is decreasing
with $Q^2$ at a slope steeper than pQCD scaling.
Exclusive experiments~\cite{Frolov:1998pw,Joo:2001tw,Ungaro:2006df} 
unambiguously show that one \emph{has not} reached a 
kinematic region where pQCD contributions become dominant up to a momentum transfer of
almost $Q^2=6$~\ufourmomts.  However, it is also possible that the data
is beginning to show an interpolating behavior between the
values at the currently accessible kinematic regions and the pQCD predictions.
Some simple expectations have been put fourth based on
the knowledge of the behaviors of other known form factors and specific
pQCD predictions~\cite{Carlson:1998bg}. 

The goal of this experiment 
was to measure  the $N \to \Delta$  transition  form factors at the highest possible 
momentum transfers and to  confront current theoretical issues:
\begin{itemize}
\item[$\bullet$]Whether $G_M^*$ continues to fall anomalously fast as a function of $Q^2$, or whether it begins to approach the scaling behavior equivalent to the dipole form. 
\item[$\bullet$]Whether E2/M1 remains very small and negative, or whether it begins to
turn positive, and asymptotically begin to approach +1.

\item[$\bullet$]Whether  S1/M1 also approaches a scaling behavior, constant with $Q^2$.  

\end{itemize}

The data presented here will facilitate the examination of the $N \to \Delta$ 
amplitudes vis-a-vis  the prediction of theoretical formalisms  in this 
higher $Q^2$ but sub-pQCD kinematic region.

The new measurements 
reported here are for the reaction $\gamma^{\ast}+p\to  \Delta^+\to p^{\prime}+\pi^0$.  Previous 
experiments at Jefferson Lab for this reaction ~\cite{Joo:2001tw,Frolov:1998pw,Ungaro:2006df} have provided data up
$Q^2$ = 6.0~\ufourmomts. The present experiment provides data of higher statistical accuracy 
at  $Q^2$ = 6.4 and 7.7~\ufourmomts, which was the highest possible at the   
beam energy of  5.5~\uenergy.  In the future, the Jefferson Lab upgrade, will enable the experiments to approach $Q^2$ values near 13 or 14~\ufourmomts.

\section{\label{S:piprod}Electroproduction of $\pi^0$ Mesons}
The single dynamical assumption which is made that makes
kinematics simpler and indeed even allows straightforward parameterization of 
the dynamics is the \emph{single photon perturbative approximation}.  The 
results of this work relating to dynamical form factors are valid only to 
the extent that this approximation is satisfied.  It is also very important
to understand the process at hand in both the laboratory and the center of
mass frames, to be defined in what follows.  This is essential because the 
measuring apparatus are understood more fully in the lab frame while the 
dynamical predictions are simplified in the center of mass frame. 

\subsection{\label{S:choord_cross}Definition of Coordinates and Cross Sections}
\begin{figure}[!htb]
\begin{center}
\includegraphics[scale=0.3]{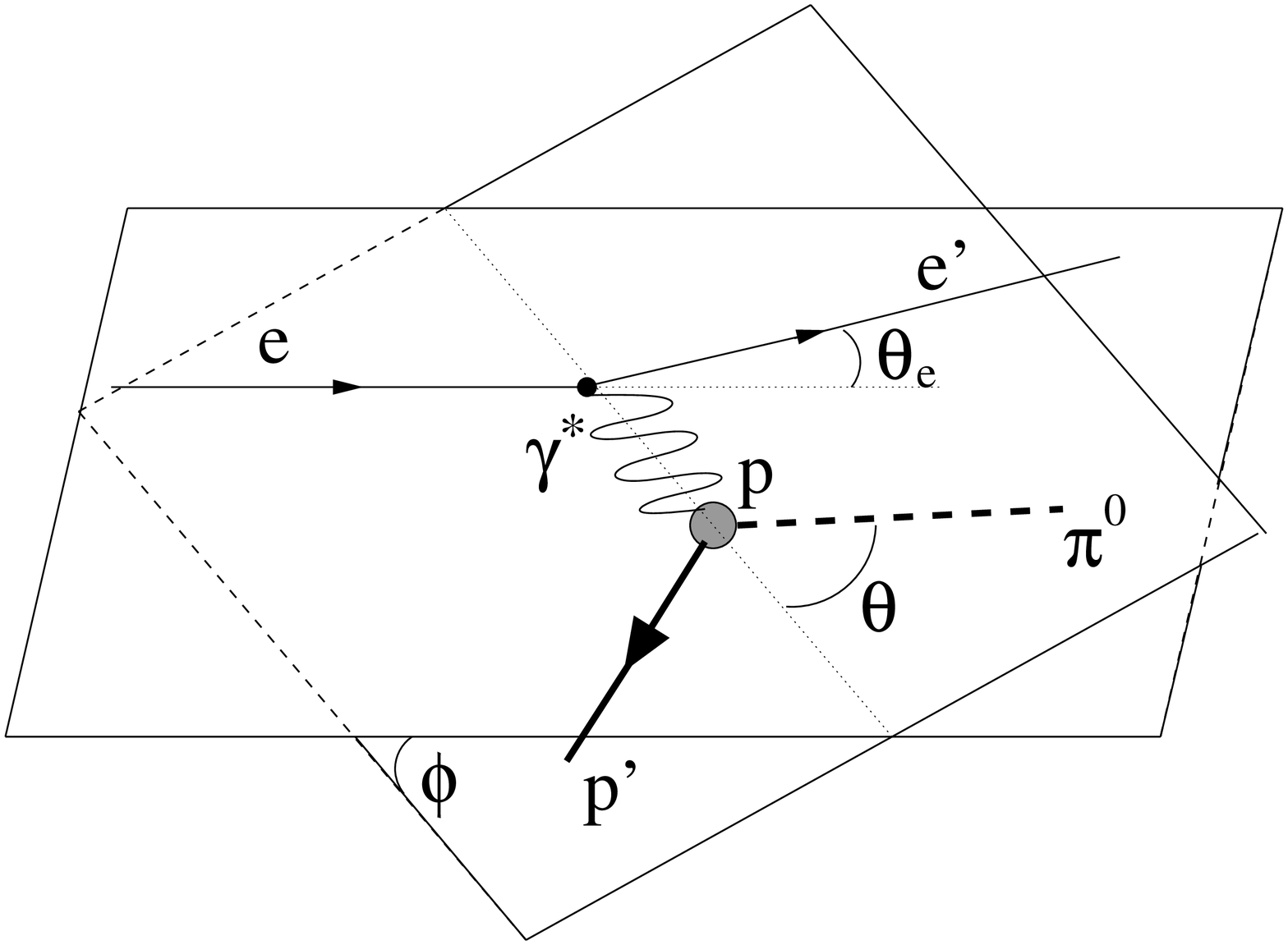}
\caption{\label{F:piprod_lab}Lab frame neutral pion production.  The symbol $e$ represents the incoming
electron and $e^{\prime}$ the outgoing electron.  The incoming and outgoing protons
are denoted by $p$ and $p^{\prime}$, respectively.  The symbol $\pi^0$ is the outgoing
neutral pion and $\gamma^{\ast}$ is the exchanged photon.}
\end{center}
\end{figure}

We examine the differential cross section for a neutral
pion from the following exclusive reaction:
\begin{equation}\label{E:exl_piprod}
e + p \rightarrow e^{\prime} + p^{\prime} + \pi^0.
\end{equation}
The kinematics for such a process are displayed in \fig
\ref{F:piprod_lab}.
In electroproduction of a single meson five kinematic variables are needed to 
specify the unpolarized reaction fully.  Assuming that the energy of the 
incident electron, $E$, is known and that the target energy is simply $m_p$, these variables can be chosen to be
the scattered electron energy $E^{\prime}$,
the electron angles $\Omega_e$ and the meson angles $\Omega_{\pi}$.
These completely specify the reaction.  Given this 
convention, the 5-fold differential cross section can be obtained as a function
of the mentioned variables.
We express as many of the variables as possible through the 
use of Lorentz invariants.
This procedure also makes one able to predict some
simple dynamical effects from the covariant procedures for calculating 
QED matrix elements (the Feynman rules).  Another advantage 
is that the lepton current portion will completely 
factorize in a frame-invariant way, which enables one to write the amplitudes
in terms of only hadronic variables multiplied with some known (and frame
invariant) QED factors.  The most obvious new coordinate which is suggested
from the lab frame kinematics 
and the canonical treatment of the elastic process is the 
momentum transfer from the electron to the target proton.
In view of the one-photon exchange approximation this can be
viewed as the 4-momentum of the exchanged virtual (off-shell) photon.
This
understanding of the 4-momentum transfer will be especially useful when moving
to the center of mass frame.
\begin{equation}\label{E:piprod_q}
q^{\mu} = k^{\mu} - k^{\prime\mu}
\end{equation}
The symbol $k$ is the 4-momentum of the incoming electron and $k^{\prime}$ 
is the 4-momentum of the outgoing electron.  Defining the incoming
proton 4-momentum to be $p$ and the outgoing to be $p^{\prime}$, the two
electron invariants become the following.
\begin{equation}\label{E:e_invariants}
\begin{aligned}
Q^2 & \equiv -q^2 = 2EE^{\prime}(1-\cos{\theta_e}) \\
W & \equiv \sqrt{(q+p)^2}=\sqrt{m_p^2 +2q^0m_p - Q^2}
\end{aligned}
\end{equation}
The rightmost equalities in \eq~\ref{E:e_invariants} hold in the lab frame. 
Another experimentally useful
invariant is the missing mass, $M_x^2$, which is the square of the undetected
4-momentum.  In the present case this is:
\begin{equation}\label{E:define_mx}
M_x^2 = (q + p - p^{\prime})^2.
\end{equation}
The dependence on the leptonic variables is now completely in terms of 
invariants which can be calculated in any frame.
\par
\begin{figure}[!htb]
\begin{center} 
\includegraphics[scale=0.3]{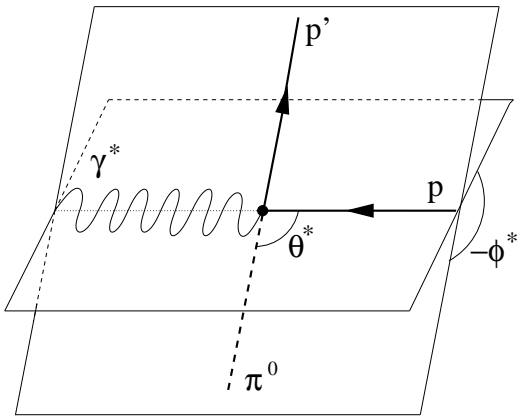}
\caption{\label{F:piprod_cm}Center of mass frame neutral pion production.}
\end{center}
\end{figure}

It is desirable to move to the hadron-virtual photon
center of mass frame.  Kinematically this is desirable because
it essentially replaces three body final state with the two 
body version.
Dynamical considerations for the pure QED portion of the matrix element 
must, however, be taken into account.  
As previously indicated, the lepton current portion of the matrix element will
factorize.
Lorentz boosting to the center of mass
along the direction of the momentum transfer enables one to
treat the hadronic cross section as the interaction of a virtual
photon with a target hadron and treat the leptonic current as a prefactor
to the amplitude which is a function of the Lorentz invariants $Q^2$ and $W$.
The center of mass frame is shown in \fig~\ref{F:piprod_cm}.
An asterisk denotes a center of mass quantity except when symbolically
referring to a photon in which case an asterisk (as in $\gamma^{\ast}$) denotes
that the photon is virtual (off-shell).
\par
The details of the lepton current factorization are
reviewed in \pref~\cite{dhL:1978ds,
Dombey:1969wk,hP:1967ti}. 
The result is that the 5-fold differential cross section can be written as follows.
\begin{equation}\label{E:mesprod_succinct}
\frac{d\sigma}{dE^{\prime}d\Omega_ed\cm{\Omega}_{\pi}} = \Gamma
\frac{d\sigma^{\gamma^{\ast}}}{d\cm{\Omega}_{\pi}}
\end{equation}
The factor $\Gamma$ in \eq~\ref{E:mesprod_succinct} is the
virtual photon flux factor.  In the Hand convention~\cite{Hand:1963bb}
this reads:
\begin{equation}\label{E:flux_define_succinct}
\begin{aligned}
\Gamma &\equiv \frac{\alpha}{2\pi^2}\frac{E^{\prime}}{E}\frac{(W^2-m_p^2)}{2m_pQ^2}\frac{1}{1-\epsilon} \\
\epsilon &\equiv \left(1+2\frac{|\mathbf{q}|^2}{Q^2}\tan^2{\frac{\theta_e}{2}}\right)^{-1}, \\
\end{aligned}
\end{equation}
in which $\epsilon$ describes the ratio of longitudinal to transverse polarization
of the virtual photons.
Because of the structure of the virtual photon
density matrix~\cite{Dombey:1969wk,aV:2007np}, one can write explicitly the $\cm{\phi}$
dependence of the center of mass cross section in terms of
the transverse (T), longitudinal (L), transverse-transverse
interference (TT) and longitudinal-transverse interference (LT)
portions of the interaction.
\begin{widetext}
\begin{equation}\label{E:diffpi_domega_succinct}
\frac{d\sigma^{\gamma^{\ast}}}{d\cm{\Omega}_{\pi}} =\sigma_T
+ \epsilon \sigma_L + \epsilon
\sigma_{TT} \cos{2\cm{\phi}} + 
\sqrt{2\epsilon(1+\epsilon)} \sigma_{LT} \cos{\cm{\phi}}
\end{equation} 
\end{widetext}
The goal of the experiment is to obtain the center of mass pion differential
cross sections
and interpret all of the
components displayed in \eq~\ref{E:diffpi_domega_succinct}
in terms of multipole amplitudes from the pion production data in this work.

\section{\label{S:exp}Experimental Overview}
The experiment was carried out in the
Jefferson Laboratory Hall C using a two-spectrometer setup for detection
of outgoing electrons and protons.
\par
\begin{figure}[!htb]
\begin{center} 
\includegraphics[scale=0.77]{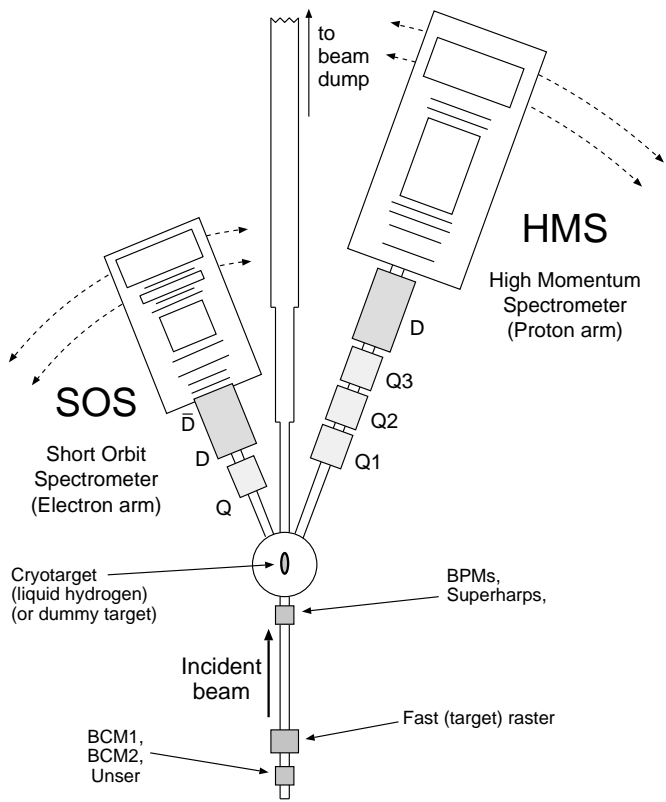}
\caption{\label{F:HC}Plan view of the experimental layout in Hall C (from
\sref~\cite{Blok:2008jy}).  The symbols Q, Q1, Q2 and Q3 denote quadrupole
magnets, D denotes forward bending dipole and $\rm \bar D$ denotes reverse bending
dipole.}
\end{center}
\end{figure}

A schematic of the Jefferson Lab Hall C setup is shown 
in \fig~\ref{F:HC}.  The hall is equipped with two
magnetic spectrometers: the High Momentum Spectrometer (HMS)
and the Short Orbit Spectrometer (SOS).
The target consisted of 
liquid hydrogen (LH$_2$), at a temperature
of 19.0 K.   
Exclusive electroproduction data for the process $ep \rightarrow epX$
was gathered in the spring of 2003 run period.  The electron
beam energy was about 5.5~\uenergy \ and the $Q^2$ values were 6.4 and 7.7~\ufourmomts \ at the
$\Delta$ resonance. 
\par
The HMS was 
used to measure the proton momentum and angles while the 
SOS was used to measure the electron momentum and angles.  Details
of the spectrometer properties and detector
packages as used in this experiment can be found in \sref~\cite{aV:2007np}.
Though the 
magnetic spectrometers have a small acceptance compared to 
the acceptance of a $4\pi$ detector, the
relatively low values of $W$ and high values of $Q^2$
cause protons to emerge in a rather narrow cone around the $\mathbf{q}$ vector. Full coverage can thus be 
obtained in the center of mass variables by
using several HMS
angle and momentum scans.  The spectrometer settings for the experiment 
are listed in \tab~\ref{T:kinmat}.
\begin{table}[!hbtp]
\begin{center}
\begin{tabular}{cc|cr}
\hline
\hline
\multicolumn{2}{c|}{Electron Arm} & \multicolumn{2}{c}{Proton Arm} \\
$p_{\mathrm{SOS}}$ & $\theta_{\mathrm{SOS}}$ & $p_{\mathrm{HMS}}$ 	& \multicolumn{1}{c}{$\theta_{\mathrm{HMS}}$}\\
\umom		& degrees		& \umom			& \multicolumn{1}{c}{degrees}		\\
\hline
	&		& 4.70\phantom{$^\dag$}		& 18.0, 15.0 \\
	&		& 4.50\phantom{$^\dag$}	& 19.5, 16.5, 13.5, 11.2 \\
	&		& 3.90\phantom{$^\dag$}  	& 21.0, 18.0, 15.0, 12.0 \\
	&		& 3.73\phantom{$^\dag$}	& 22.5, 19.5, 16.5, 13.5, 11.2\\
1.74 	& 47.5 		& 3.24\phantom{$^\dag$} 		& 24.0, 21.0, 18.0, 15.0, 12.0\\
	& 		& 3.10\phantom{$^\dag$}	& 22.5, 19.5, 16.5, 13.5, 11.2\\
	&		& 2.69\phantom{$^\dag$}		& 24.0, 21.0, 18.0, 15.0, 12.0\\
	&		& 2.57\phantom{$^\dag$}  	& 22.5, 19.5, 16.5, 13.5, 11.2\\
	&		& 2.23\phantom{$^\dag$}		& 21.0, 18.0, 15.0, 12.0 \\
	&		& 2.13\phantom{$^\dag$}	& 22.5, 19.5, 16.5, 13.5 \\
\hline
	&		& 4.70 & 11.2 \\
 	&		& 4.50 & 14.2 \\
1.04 	& 70.0 		& 3.90 & 11.2 \\
 	&		& 3.73 & 14.2, 11.2 \\
 	&		& 3.24 & 11.2 \\
\hline
\hline
\end{tabular}
\end{center}
\caption[Kinematics]{\label{T:kinmat}The kinematic settings of the two spectrometers.  The beam energy is nominally 5.5~\uenergy.}
\end{table}
\subsection{\label{S:beam_targ}Beamline and Target}
 The experiment depends on knowing to a reasonable accuracy the beam energy,
and current.  
Prior to the interaction in the target the electron beam traverses the beam current monitoring, beam
energy measurement and beam raster devices. 
\par
In standard running, the beam is tuned in an achromatic mode through an arc
which consists of eight dipoles and is located just before the beam enters Hall~C.  
To measure the beam energy, the beam is  tuned to a dispersive mode 
through the arc dipoles. The current in the 
arc dipole magnets is varied until the beam is centered at the exit of the 
dipole arc. The relationship between the current in the arc dipoles and the field integral
is known from previous measurements. The angle and position of the beam when entering and exiting the arc
are measured and used to determine the correct path length through the arc dipoles.
The relative  uncertainty on the beam energy measurement is 5$\times$10$^{-4}$ which is due to
uncertainty in the field integral and in the path length through the arc dipoles.
\sref~\cite{Yan:1993} is a detailed description of the beam energy measurement
technique. The beam energy measurement was done only once during the experiment, since
the measurement interrupts regular data taking. To monitor changes in the
beam energy during the experiment relative to the arc energy measurement, the 
positions and angles of the beam in the arc dipoles are measured throughout experiment
and the beam energy is determined continuously.
The beam energy varied during the first quarter of the experiment.  
The beam energy varied from 5.501~\uenergy \ to as low as 5.492~\uenergy.  
After this period, the beam energy was stable at 5.499~\uenergy.
The small beam energy difference was taken into account in all simulation work 
and data reconstruction.  Since results are not reported as a function
of beam energy and the values of kinematics were calculated with the appropriate
value for $E$, the beam energy is stated to be 5.5~\uenergy \ throughout
this work when listing kinematics.
\par
The beam current measurement is accomplished by using two beam current
monitors (BCMs) positioned along the beam line.  These current monitors
are quite stable but do not have the ability to make an accurate absolute
measurement.  An additional current monitor, the Unser monitor, has a
very stable gain
but an offset that drifts considerably on short time scales~\cite{Unser:1981at},
experiencing typical drifts of 3~$\mu$A.
The solution used in this experiment was
to extract the Unser monitor zero at various intervals during the experiment
by ramping the beam current down in several steps.
The BCMs, which are more stable but lack the absolute accuracy of the Unser,
are then calibrated with the Unser monitor.  This method was measured to be stable
to 0.2\% from run-to-run and had an overall accuracy of 0.5\% on the charge
measurement~\cite{Blok:2008jy}.
\par
After several current monitors on the beam line there is the fast raster system~\cite{Yan:1995uu}. 
The Jefferson lab electron beam has very small spacial extent
and therefore would induce significant boiling in cryogenic targets if the
beam were allowed to impinge on the target for too long at a current of
a few to several tens of microamps.  For this reason, Hall C uses the fast raster 
which sweeps the beam uniformly over a square pattern on the target.  The
size of this pattern is typically $\pm$1.2 mm in the horizontal and vertical directions.
\par
It should also be noted that the beam itself has a periodic time
structure due to the RF techniques used to create and accelerate the beam.  For the
Jefferson Lab accelerator the frequency of this structure (corresponding
to the excitation frequency of the cryogenic accelerator cavities) is 
1497~MHz which corresponds to beam pulses which are about 668~ps apart.
The beam is delivered to each hall by a kicker magnet which moves a third of the beam
into each of the three hall beam pipes.  Therefore when the beam arrives in
each hall it will have bunches which are separated by roughly 2~ns.  This
intrinsic beam structure was important for subtracting coincidence
spectrometer events which have two particles that \emph{do not} correlate
to the same beam bunch.   
\par
We turn now to the target specifications.
The geometry of the target is especially important because of the
possibility of electron scattering interactions in the target walls.
The LH$_2$ target
was kept in a constant cooling loop with a temperature
of 19.0~K and pressure of 24~psi.  At this temperature and pressure, the density
of liquid hydrogen is 0.0723~g/cm$^3$.
The target ladder for the experiment contained several other
targets along with a ``dummy'' target which was used for measuring
the contribution to the data due to scattering in the target walls. 
This experiment used the LH$_2$ target
and the Al dummy only.
The target cell was cylindrical and 4.013$\pm$0.008~cm in diameter,
made of 7075 aluminum with the beam impinging on the non-circular face.
The thickness of the target cell was measured at four places around
the cylinder~\cite{Meekins:2003tc} and the results average to 0.1330 $\pm$ 0.0013~mm.
There was a beam offset of 3~mm from the center of the cell so that the active length of the
target included 
3.941~cm of liquid. 
Electron radiation from this material was included in the Monte Carlo
simulation used for the data analysis.
\subsection{\label{S:detlog}Detector Properties}
\begin{figure}[!htb]
\begin{center} 
\includegraphics[scale=0.3]{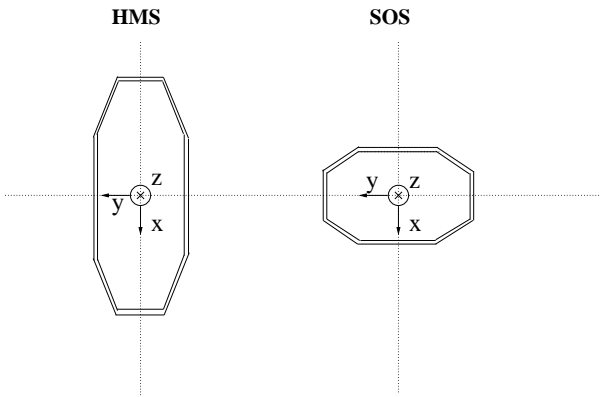}
\caption{\label{F:spec_axis}Spectrometer coordinate systems.  The octagonal shapes represent the boundaries
of the collimators at the entrances of the first quadrupole magnets of each spectrometer.}
\end{center}
\end{figure}

The spectrometer coordinate system is defined such that the ``z'' axis is along the central axis of the 
spectrometer, ``x'' axis points in the positive 
dispersive direction and the ``y'' is perpendicular to the dispersive plane defined by the choice
of a right handed coordinate system.  Figure~\ref{F:spec_axis} shows 
the coordinate systems of both the SOS and HMS spectrometers.  Both the 
focal plane and target quantities use this coordinate system for 
detected particles.  In particular the change in the x or y coordinates 
per unit change in the z coordinate is used to calculate angles.   
The entrances to the spectrometers are equipped with collimators having 
different dimensions for the HMS and the SOS.  The octagonal shape of the collimators
are displayed in \fig~\ref{F:spec_axis} centered around the coordinate
axes.
The flight distance from the target to the collimator is 
166.4~cm for the HMS and 126.3~cm for the SOS spectrometer.  Each of the 
collimators are 6.3~cm thick and with beveled interiors so that
exit openings are slightly larger
than the entrance openings. 
\par
The HMS and SOS use momentum dispersion due to dipole magnetic fields in
order to analyze the momentum of particles.  Different momenta will pass 
through at different positions inside the detector hut on a two dimensional
surface referred to as the focal plane. 
\par
The magnet configuration for the HMS is $QQQD$ (three quadrupoles then a dipole)
and the configuration for the SOS spectrometer is $QD\bar{D}$, where the bar 
denotes a central bend angle in the opposite direction.  The quadrupoles are used as
focusing elements in general to allow the apparatus to accept events which would
hit the spectrometer material had they not been focused prior to bending
\cite{Green:2000tp}.  Both the HMS and SOS spectrometers used a
point-to-point magnetic configuration, wherein particles which originate
from a common point with common momenta will be focused to the same point
on the focal plane.  The magnets in the spectrometer are typically modeled
by transport matrices in phase space where the matrix elements are fitted
to data or obtained from a precise field map.  Procedures for the optimization
of the matrix elements for the magnets in Hall C have been refined over
the years~\cite{dDutta:1996hc,Blok:2008jy}.  The SOS dipole magnet
saturates above about 1~\umom \ in momentum so that a separate transport matrix had
to be used for the 1.74~\umom \ (low $Q^2$) setting in the current experiment.
The HMS had the same magnet matrix for all settings.  This fact leads to
a somewhat poorer knowledge of the SOS acceptance than the HMS acceptance
which can be checked by measuring inclusive data in each spectrometer.  The
SOS acceptance was studied by using inclusive electron scattering and
results are presented in \sref~\cite{Dalton:2008ff}.  The HMS acceptance
has been extensively studied in electron inclusive scattering
experiments~\cite{Christy:2004rc,Tvaskis:2006tv}.
\par
\begin{figure}[!htb]
\begin{center} 
\rotatebox{-90}{\includegraphics[scale=0.35]{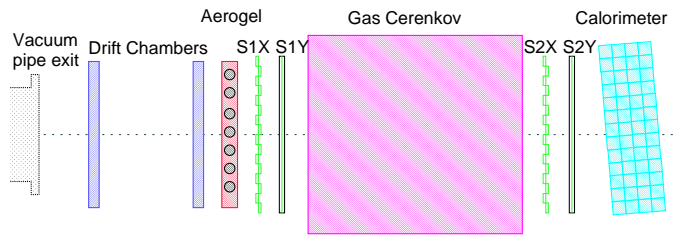}}
\caption{\label{F:det_pack}(color online) Typical spectrometer detector package.  The graphic was
taken from \sref~\cite{Blok:2008jy}.  Particles travel from left to right.}
\end{center}
\end{figure}

Figure~\ref{F:det_pack}
shows the typical detector package which is utilized in each spectrometer hut.
Drift chambers are located on either side of
the focal plane in each spectrometer and are shown schematically in \fig~\ref{F:det_pack}.
The drift chambers are used to determine the detected particle's position and
direction in each spectrometer's focal plane.  The rest of the detector
package is located after the last drift chamber.  
\par
The two sets of X-Y hodoscopes are shown on either
side of the gas \v{C}erenkov detector. These are labeled S1X, S1Y, S2X and S2Y in order along
$+\hat{\mathbf{z}}$, with the X or Y label referring to the orientation of the scintillator strips.
The hodoscopes were used for the electronics trigger in
a 3-out-of-4 configuration, that is, a pretrigger is generated if 3 out of
4 of the hodoscope planes fire. 
\par
\begin{figure}[!htb]
\begin{center} 
\includegraphics[scale=0.32]{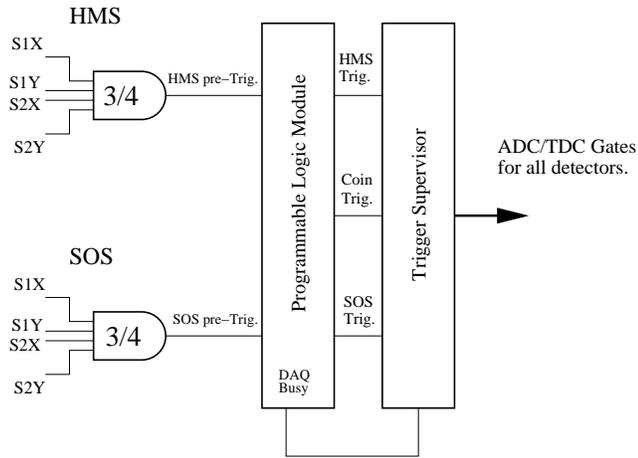}
\caption{\label{F:daq_elec}Simplified representation of the electronics and data acquisition
system for experiment E01-002.
}
\end{center}
\end{figure}

The basic electronics selection mechanisms and read out scheme is represented
in \fig~\ref{F:daq_elec}.
The scintillator bars on the four hodoscope X or Y planes were read out at
each end and used to create a pretrigger.  A signal on either edge of the bars give
an electronic logical true if any of these bars fire.
As \fig~\ref{F:daq_elec} indicates, these pretriggers were then
passed to a programmable module which decides which kind(s) of data acquisition
triggers to produce.  The so-called ``8LM'' programmable module will \emph{not}
produce a data acquisition (DAQ) trigger if it receives a
``busy'' signal from the DAQ, indicating that the DAQ is not ready for another
event.  When the DAQ was not busy, the 8LM module produced HMS, SOS,
or coincidence triggers which were passed along to the \emph{trigger supervisor}.  The
coincidence trigger was the logical ``and'' of the HMS and SOS triggers
which require a 3 out of 4 scintillator plane event
in \emph{each} spectrometer.  The timing between the SOS and HMS pretriggers
was adjusted so that there was an overlap
for a coincidence trigger.
\par
The trigger supervisor controls
the DAQ by dispensing
gates to the ADC and TDC modules only when a valid event is present
\emph{and} the DAQ is not already busy digitizing a previous event.
The trigger supervisor also performed any necessary prescaling of the signals.
The prescaling allows the DAQ to skip some set number of
events, or, read only every $n^{th}$ event where $n$ is the prescale factor.
For example SOS singles prescale factors used in this work  were 1, 2, 3 and 5.
The
HMS singles have prescale factors which ranged from 100 to a few
thousand at the low hadron momentum settings where $\pi^+$ production
is copious.
After
the appropriate gates are dispensed for the appropriate triggers, the ADC
and TDC modules (located in FastBus crates) will begin to digitize all relevant
information concerning analog photomultiplier signals and time difference
signals.
\par
After being triggered the computer DAQ system digitized and stored
the information from all of the detectors and monitors. 
The Jefferson Lab CODA
\cite{Abbott:2004chep} event
builder was used to retrieve all relevant information from
the ADC and TDC modules while storing event information on disk and/or
on tape.  Internet connections were used to communicate with
the CPUs which were storing the ADC or TDC results.

Several data restrictions were made simply to ensure that the
analyzed events include only ones where the
SOS spectrometer recorded an electron event, the HMS spectrometer
recorded a proton event and that these events are in coincidence. 
\par
Before the particle identification selections were made, however,
the so-called ``fiducial volume'' was restricted to
ensure that we use parts of the spectrometer focal plane which are
well understood and avoid optics ambiguities.  This allows the
acceptance to be well modeled by Monte Carlo techniques. 
The fiducial
restrictions were:
\begin{equation}\label{E:fid_excl}
\begin{aligned}
 -20.0 \le &\; X_s \le 22.0 \quad  (\text{mm}) \\
y^{\prime}_{min} \le &\; y^{\prime}_s \le y^{\prime}_{max} \\
-18.0 \le &\; \delta_s \le 18.0 \quad  (\%)\\
-9.0 \le &\;\delta_h \le 9.0 \quad (\%).
\end{aligned}
\end{equation}
\par
\noindent The symbols $y^{\prime}_{min}$ and $y^{\prime}_{max}$ are  defined as follows.
\begin{equation*}
\begin{aligned}
y^{\prime}_{min} &\equiv \frac{1}{1000}(-125.0 + 4.25\delta_s + 64.0 y_s -1.7\delta_s y_s) \\
y^{\prime}_{max} &\equiv \frac{1}{1000}(125.0 - 4.25\delta_s + 64.0 y_s -1.7\delta_s y_s) \\
\end{aligned}
\end{equation*}
\par
The symbol $\delta_s$ is defined as $(p-p_c)/p$ with $p_c$ the central momentum in
the SOS spectrometer and $p$ the detected particle momentum, $\delta_h$ is the analogous quantity for the HMS,
$y_s$ is the SOS ``y'' position at the target, $X_s$ is the SOS ``x''
position at the focal plane, and $y^{\prime}_s$ is the SOS ``y'' angle
at the target.  A further fiducial restriction is made by removing events
which reconstruct to outside either of the collimator apertures.
\par
\begin{figure}[ht]
\begin{center} 
\includegraphics[scale=0.47]{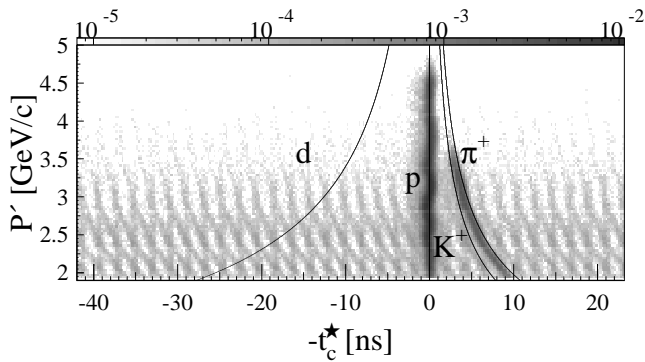}
\caption{\label{F:cointime_prot_norm}The ``relative'' $t^{\star}_c$ vs. hadron momentum for coincidence data 
normalized to make the proton's time constant for all proton momenta.} 
\end{center}
\end{figure}

The particle identification restrictions include two restrictions
to identify electrons in the SOS spectrometer along with a timing
restriction to verify that the HMS detects a coincident proton.
Figure~\ref{F:cointime_prot_norm} displays the HMS momentum
vs. the corrected time signal called ``coincidence time,'' $t^{\star}_c$.
The corrected time signal is constructed so that the proton events
arrive at zero relative time independent of momentum. The figure displays
the relative timing curves for other possible HMS contaminants.  One can see
that the $\pi^+$ signal is largest but still easily separated from the proton
signal.
The particle identification 
requirements are listed in \eq~\ref{E:cointime_excl} where
the coincidence time is relative to the center of the proton peak.
\begin{equation}\label{E:cointime_excl}
\begin{aligned}
|t^{\star}_c| &\le 1.5 \quad  (\text{ns}) \\
\epsilon_s & \ge 0.8 \\
N^{\gamma}_s &\ge 0.5 
\end{aligned}
\end{equation}
The variable $\epsilon_s$ represents the energy deposited in the SOS calorimeter divided
by the particle momentum.  The symbol $N^{\gamma}_s$ is the number
of \v{C}erenkov photons detected in the SOS.
\par
\subsection{\label{S:W_mmoverall}Data Overview }
\begin{figure}[!htb]
\begin{center}
\includegraphics[scale=0.45]{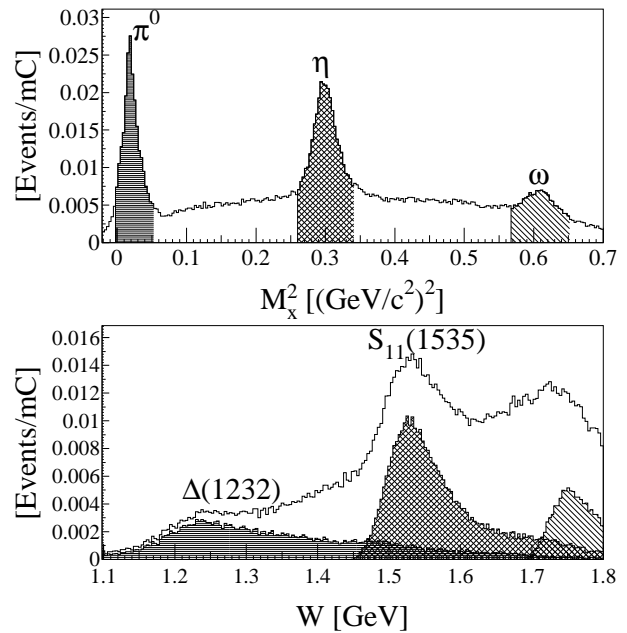}
\caption{\label{F:mmass}Upper panel: the missing mass distribution.
The shaded regions highlight the peaks corresponding to
$\pi^0$, $\eta$ and $\omega$ production.  Lower panel: the
$W$ distribution corresponding to the events in the upper panel.  the shaded regions in the
lower panel correspond to events in the shaded region of the
upper panel.  Each of the plots report the number of counts normalized
to the total charge in millicoulombs~(mC).}
\end{center}
\end{figure}

It is useful to examine the overall results of the experiment to obtain
intuition about backgrounds and cuts.  The most natural distributions 
to look at are the missing
mass, $M_x^2$, and invariant mass, $W$, distributions in \fig~\ref{F:mmass}.
The invariant mass is that of the
virtual photon-nucleon system and was quantified in \eq~\ref{E:e_invariants}.
The missing mass distribution shows peaks corresponding to exclusive single mesons and continua due
to multi-meson production and background.  The invariant mass distribution indicates along which regions of invariant
mass the meson events come from.
Clear correlations between $W$ and
$M_x^2$ can be seen in the figure.
Further, one can see that 
$\pi^0$ production peaks at the $\Delta(1232)$ 
resonance and $\eta$ at the $S_{11}(1535)$ 
resonance.
It should be noted that the $\Delta(1232)$ resonance
is by no means the only source of $\pi^0$ production, whereas
the $S_{11}(1535)$ dominates the $\eta$ production in the present region of $W$. 
\par
The invariant mass thresholds for $\pi^0$, $\eta$ and $\omega$ production
are 1073.3, 1486.1 and 1720.9~MeV respectively.  The two pion production
threshold is at $M_x^2=$0.0729~\umasss, above the $\pi^0$ production
analysis restrictions used in the present work. 

\section{\label{S:bknd}Background Subtractions}
Not all of the events present in the raw data acquisition
represent the physical process of interest.  One therefore
must remove or modify a significant amount of events before the
analysis can proceed.  These modifications of the data set come
in several varieties including background subtraction,
and data
corrections.  Each of these modifications
are considered in turn with specific attention given to
radiative corrections on the pion production amplitudes, which
have a physical origin distinct from any detector effects.

\subsection{\label{S:physback}Radiative Background Processes}
There are two types of radiative processes which need to be treated.
Radiative elastic scattering gives a background to the $\pi^0$ peak in the missing mass spectrum.
Radiative processes accompanying $\pi^0$ electroproduction deplete the number of events under the single $\pi^0$ missing mass peak. 
\par
This section concerns elastic radiative processes which may
``masquerade'' as pion electroproduction processes in data analysis.  The 
elastic radiative process is represented in \eq~\ref{E:kin_el_eqn}. 
\begin{equation}\label{E:kin_el_eqn}
e + p \rightarrow e^{\prime} + p^{\prime} + \gamma
\end{equation}
The $\pi^0$ electroproduction process is:
\begin{equation}\label{E:kin_pion_eqn}
e + p \rightarrow e^{\prime} + p^{\prime} + \pi^0
\end{equation}
and is, in principle, easily distinguished from the radiative
process but because of finite detector resolutions care must be
taken in separating the two.  
\par
The missing mass for this work is always calculated by summing 
the 4-momenta of the incoming and outgoing measured particles.  With the
standard kinematic conventions one has that $p_m^2=(k+p-k^{\prime}-p^{\prime})^2=M_x^2$.
For elastic scattering, or the case where a single photon is radiated, $M_x^2=0$.
The low mass of the $\pi^0$, $m_{\pi}^2\sim$ 0.018~\umasss, makes it difficult to 
separate from processes which have $M_x^2=0$.  This is because 
of experimental resolution effects on the calculated missing momenta.
The result
is that
the pion and radiative missing mass peaks will have an apparent broadening and, depending on $Q^2$
and $W$, the peaks may overlap.  Generally speaking, the widths of the $M_x^2$ peaks are
smallest for $W$ near the elastic peak and become larger with increasing $W$, so that in the
region at or above the peak of the $\Delta(1232)$ there is a significant overlap
of the $\pi^0$ and elastic missing mass peaks.   
\par
The radiative processes of QED have been studied for many years and an authoritative
body of literature exists on the subject
\cite{Maximon:2000hm,jdWAL01,Ent:2001hm,Mo:1968cg}.  Some of the major
developments were the treatment of the infrared divergences and the re-summing
of the QED expansion for multiple low-energy photons 
\cite{Yennie:1957}.  In this work the resulting angular and energy dependences 
of the radiative events are used to remove elastic radiative contamination from
the pion production peak. 
\par
The amplitudes for initial or final state radiation can be calculated
exactly using well known QED techniques and suitable parameterizations
of proton elastic form factors~\cite{mROS50}.
An immediate result of the photon radiation amplitudes is that
there are strong peaks along the direction of the outgoing or incoming
charged particles.  Since the proton is about two thousand
times more massive than the electron, the
radiation will be predominantly along the directions of the
incoming and outgoing electrons.  This fact is an important
kinematic reality that allows this contribution to be
excluded fairly efficiently even without simulation of the
radiative events. 
\par
\begin{figure}[!htb]
\begin{center} 
\includegraphics[scale=0.47]{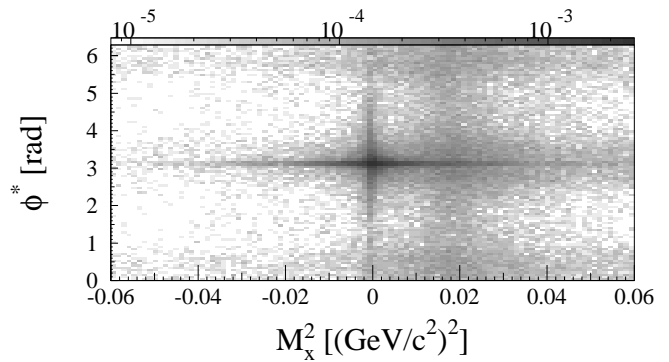}
\caption{\label{F:phimm_nocut}$M_x^2$ vs. $\cm{\phi}$ with only standard cuts and $W >$ 1.1 GeV applied.
Events in the
vertical band near $M_x^2=$ 0.018~\umasss \ are from single $\pi^0$ electroproduction. Elastic
events form a peak around $M_x^2=$ 0 and elastic radiative events form a broad band
centered on $\cm{\phi}=\pi$.
} 
\end{center}
\end{figure}

The result of this tight angular distribution is that
elastic radiative events, though they might have a recorded
invariant energy in the $\Delta$ region will emerge 
very nearly in the electron scattering plane.
In other words they will peak around
$\cm{\phi}$= $\pi$.
In contrast, the plane of emitted protons and pions can be distributed
around the electron scattering plane with $\delta \cm{\phi} \sim 2\pi$.  By
cuts close to $\cm{\phi}$= $\pi$ one eliminates nearly all the elastic radiated events
while losing only a small fraction of non-radiated events. 
The binning used for the data is such that
removing events in a tight angular region around $\cm{\phi}$= $\pi$
will not have an adverse affect on the data quality. 
The two dimensional distribution displayed in \fig~\ref{F:phimm_nocut}
shows the elastic radiative events around zero missing mass spreading
to lower and higher missing mass in a narrow line along $\cm{\phi}$= $\pi$.
A composition of
two exponential contours were chosen to eliminate the unwanted radiative
events while impacting the signal events minimally.  For this purpose
an envelope equation $\tilde{\phi}(M_x^2)$ is defined with several adjustable
parameters.
\begin{equation}\label{E:mmphicut}
|\tilde{\phi}(M_x^2)| =
\begin{cases}
\left(\frac{\phi_g}{e^{-\gamma^{\prime}m_g}}\right) e^{-\gamma^{\prime}M_x^2}+\pi, & M_x^2 < m_t \\
\left(\frac{\phi_t}{e^{-\gamma m_t}}\right)e^{-\gamma M_x^2}+\pi, & M_x^2 \ge m_t
\end{cases}
\end{equation}
With the definition:
\begin{equation}\label{E:gamprime}
\gamma^{\prime} \equiv \left(\frac{1}{m_t}\ln{\frac{\phi_g}{\phi_t}}\right).
\end{equation}
\par
The missing mass resolution becomes poorer with increasing $W$ since the range
of protons are emitted with a greater variation in momenta over a greater
range of $\cos{\cm{\theta}}$, and detected over a larger range of the
spectrometer focal plane.
Thus, the parameters which define the elastic radiative rejection
should be functions of the azimuthal angle.
\begin{table}[!hbt]
  \begin{center}
   \begin{tabular}{  c | c | c | c | c | c }
   \hline
   \hline
$\cos{\cm{\theta}}$ range  &  $\phi_t$  & $\phi_g$ & $\gamma$ & $m_t$ (GeV$^2$)& $m_g$ (GeV$^2$)  \\ \hline
-1.0 $\le \cos{\cm{\theta}} <$ -0.4  &  0.4  & 3.0 & 30.0 & 0.006 & 0.0  \\ 
-0.4 $\le \cos{\cm{\theta}} <$ 0.25  &  0.19  & .20 & 20.0 & 0.006 & 0.0  \\ 
0.25 $\le \cos{\cm{\theta}} <$ 1.0  &  0.19  & .20 & 20.0 & 0.006 & 0.0  \\ 
   \hline
   \hline
    \end{tabular}
    \caption{\label{T:thet_dep_cutpar}Radiative rejection in different $\cos{\cm{\theta}}$ bins.}
   \end{center}
\end{table} 
Table~\ref{T:thet_dep_cutpar} displays the elastic radiative rejection parameters
in each region of the azimuthal angle.  The binning in the table was chosen empirically in order
to reflect the variation in $\cm{\phi}$ vs. $M_x^2$ of the radiative tail distribution.
\par
\begin{figure}[!htb]
\begin{center} 
\includegraphics[scale=0.53]{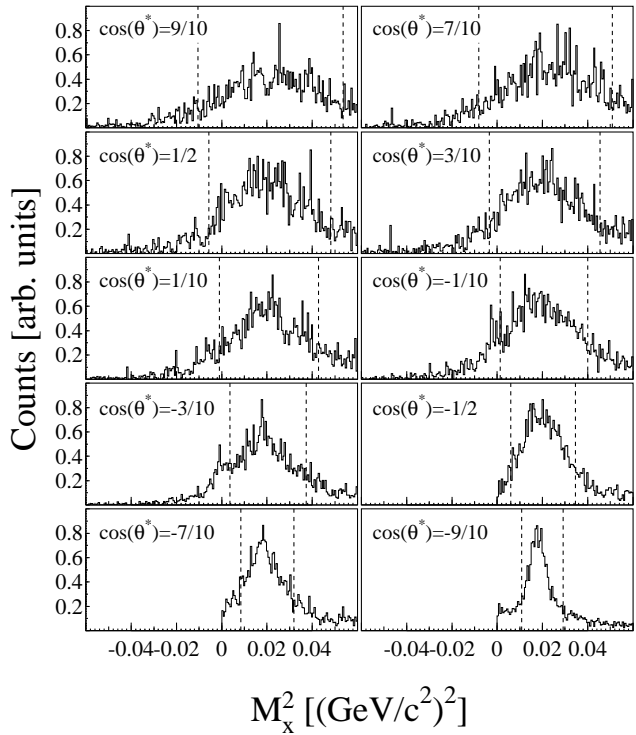}
\caption{\label{F:mm2_stdcut}Experimental missing mass ($M_X^2$) distributions for ten $\cos{\cm{\theta}}$ bins after the cuts in \tab~\ref{T:thet_dep_cutpar} were applied.   The
histograms are arbitrarily normalized and include \emph{all} $\cm{\phi}$ values. The dashed vertical lines represent the regions around the $\pi^0$ peaks outside of which the events are rejected by the 
further $\cos{\cm{\theta}}$ cuts of \eq~\ref{E:MM_excl}.  The vertical scale of the plots displays the number of events normalized to the maximum bin content of each histogram.
} 
\end{center}
\end{figure}

The two dimensional radiative rejection depends on the missing
mass ($M_x^2$).  In addition to this two dimensional restriction
there is a simpler missing mass restriction that should be applied
for the final analysis to be sure that only pion production events
are selected.
The missing mass requirement is a standard one dimensional
restriction, made with a width that is a function of 
$\cos{\cm{\theta}}$ to account for the resolution
change in the double arm measurement.  The value of
$\cos{\cm{\theta}}$ is taken to be at the center of the kinematic
bin.  The specific form
is determined by an empirical fit to the missing mass
widths.
\begin{equation*}
\begin{aligned}
M^2_{min}&=(-0.0118\cos{\cm{\theta}} + 0.00014)\\
M^2_{max}&=(0.0136\cos{\cm{\theta}} + 0.04134)
\end{aligned}
\end{equation*}
The missing mass requirement can then be expressed as the following: 
\begin{equation}\label{E:MM_excl}
M^2_{min} \le M^2_x \le M^2_{max}. 
\end{equation}
Figure~\ref{F:mm2_stdcut} displays this missing mass requirement for
several kinematic bins.
\par
In summary, the radiative elastic events have been removed from the current
data set via a restriction on the kinematic variables.  Since the
missing mass spectra look quite clean after the subtraction, no further
subtractions were needed for this background.
\subsection{\label{S:epg_prc}Background Simulation for $p(e,e^{\prime}p)\gamma$}
\begin{figure*}[!htb]
\begin{center} 
\rotatebox{90}{\includegraphics[scale=0.73]{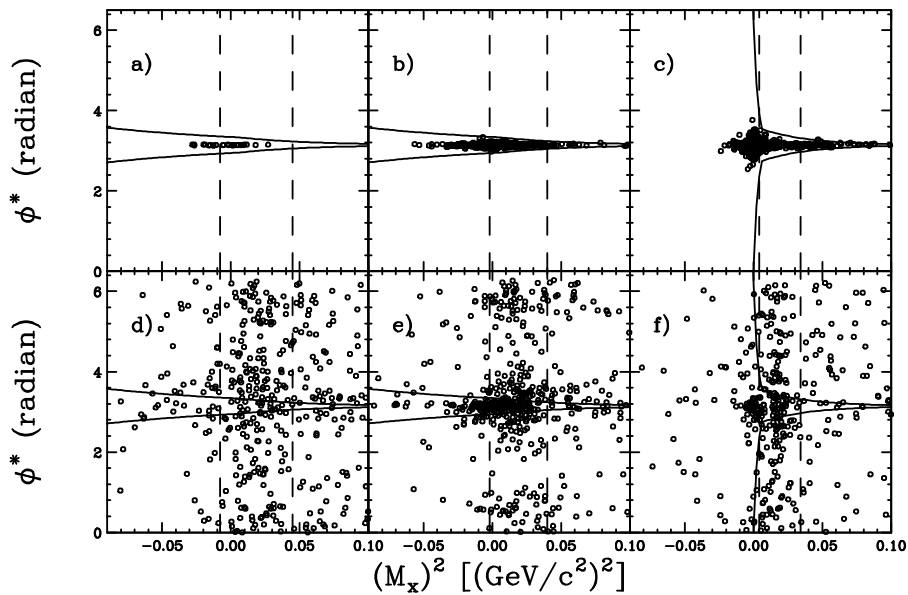}}
\caption{\label{F:epg}Comparison of the radiative rejections with simulated elastic
radiative events.  The upper panels a), b) and c) are the simulated elastic radiative
events with the cuts from \sect~\ref{S:physback} superimposed.  The lower
panels d), e) and f) repeat the data distributions for comparison. 
Panels a) and d) display the rejection for the 0.25 $\le \cos{\cm{\theta}} \le$ 1.0
range.  The panels b) and e) show the rejection curve for the -0.4 $\le \cos{\cm{\theta}} \le$
0.25 range.  The panels c) and f) show the rejection curve for the -1.0 $\le \cos{\cm{\theta}} \le$
-0.4 range.  The dotted lines represent the additional $M_x^2$ restriction placed on the
data in the analysis.
} 
\end{center}
\end{figure*}

We simulated the $p(e,e^{\prime}p)\gamma$ process with a Monte
Carlo method similar to that for the exclusive pion production.
The angle peaking approximation was used to generate photons
along the direction of incident or scattered electron (or both)
with a probability distribution based on the formulas of 
\sref~\cite{Mo:1968cg}. The elementary cross section was modeled
using the form factor parameterization of \sref~\cite{Bosted:1994tm}. 
The number of events below pion threshold $W<1.08$~\uenergy \ was found
to be in good agreement with those observed in this experiment.
The distribution of events for $W>1.1$~\uenergy \ is plotted in
three bins of $\cos{\cm{\theta}}$ in the upper panels of 
\fig~\ref{F:epg} as  a function of $M_x^2$ and $\cm{\phi}$. 
The distributions are strongly peaked near $\cm{\phi}=\pi$, as
expected, and for forward angle protons, a strong peak is also
evident near $M_x^2=0$. The curves on the plots show the
cuts used to reduce the background from events near $\cm{\phi}=\pi$
to a negligible level (less than 1\% contamination of the 
$\pi^0$ sample, in the worst case). The functional form of the
$M_x^2$-dependent cuts on $\cm{\phi}$ was described
in the previous section and the parameters were listed
in \tab~\ref{T:thet_dep_cutpar}.
The vertical dashed lines show the cuts used to remove the
events near $M_x^2=0$ (and also to reduce the background
from accidental coincidences). 
The actual distributions of events from this experiment versus
$M_x^2$ and $\cm{\phi}$ are repeated for comparison in the lower panels of 
 \fig~\ref{F:epg}. It can be seen that distributions very
similar to those in the upper panels (from the simulation) are
super-imposed on a flatter distribution from $\pi^0$ production,
and a very flat distribution from accidental coincidences. 
It was checked that the magnitude of the simulated 
$p(e,e^{\prime}p)\gamma$ background was within 20\% to 30\%
of the observed distributions. Since the background is so
concentrated in a narrow region of $\cm{\phi}$, it was decided to
not subtract this background, but simply reduce it to a 
negligible level with the cuts described above. As a further
check that the simulation matched the $M_x$ and $\cm{\phi}$ 
resolutions of the experiments, the cuts were varied over
a reasonable range, and no significant change in the cross
section was observed. This is described further in Section~\ref{S:sys}

\subsection{\label{S:datcorr}Data Corrections}
There are several corrections that must be made to the data
that are unrelated to competing physical processes but
are a result of the apparatus used for the measurement.  For the
current measurement these include ``accidental'' coincidence counts,
missed counts due to inefficiency in the data collection
process, particle tracking inefficiencies and proton absorption.
Still other effects are observed
to be small and so they are not explicitly corrected for but
are included in the systematic error estimation.  These include
target boiling, target window scattering, and calibration
inefficiencies. 
\par
\begin{figure}[!htb]
\begin{center} 
\includegraphics[scale=0.47]{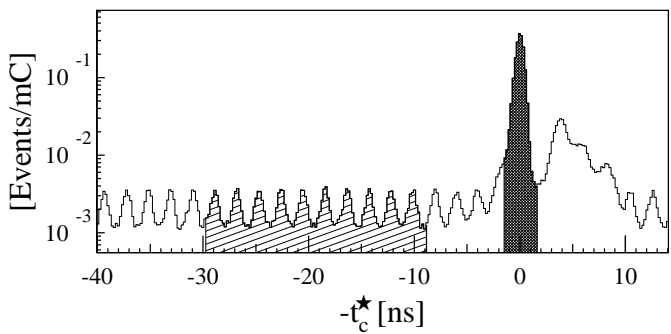}
\caption{\label{F:cointime_spec}Timing spectrum of coincidence events. The cross-hatched 
insert shows the typical analysis cut made on this spectrum.  The
diagonal shading shows typical RF beam structure populated with 
accidental coincidences.  Proton coincidence events
are normalized to appear at zero with lighter particle coincidences appearing
to the right. 
}
\end{center}
\end{figure}

Because a radio frequency (RF) pulsed beam has a characteristic timing
structure, there is a possibility that a coincidence trigger
can originate from two particles from different beam bunches.
The electron beam at Jefferson Lab's continuous electron
beam accelerator has a regular periodic structure in time and this structure
helps identify contamination from non-vertex electrons or hadrons. The 
``coincidence time'' is a variable which measures when the HMS detector 
triggers (proton detection) with respect to when the SOS detector triggered. 
\fig~\ref{F:cointime_spec} shows the distribution of events in this 
timing variable.   
\par
In \fig~\ref{F:cointime_spec}, the large peak corresponds to
proton coincident events and the large shoulder at higher
coincidence time are $\pi^+$ events (there are enough $\pi^+$
events to analyze the charged pion production cross section).  Outside
of these peaks the periodic structures correspond to events which
make a coincidence between two different RF bunches.  These events
are the ``accidental'' coincidences and are removed by using data
from far out on the coincidence time spectrum and assuming that
the structure persists through the proton peak.  The events used
for the subtraction were taken from the diagonally shaded region
in \fig~\ref{F:cointime_spec} and were far from the region where one
could expect deuteron coincidences.
\par
A more sophisticated extraction of the underlying beam 
structure might have been warranted if there were higher rates for these 
other positively charged hadrons.  The accidental count subtraction
was typically small, $< 1.0$\%, with the worst case being 8.0\% which
occurred in only one kinematic bin. 
\par
Although the timing selections select events in the SOS and
HMS spectrometers that are coincident, the electronics and computers which
allow this data to be recorded have associated deadtimes.  The resulting
computer and electronic deadtimes have been quantified. 
\par
The electronic deadtime was measured on a run by run basis
using a scalar read out of different gate generators
which are triggered by the coincidence event pulse.  Having a rate
measurement for several different generated gate widths
allows extrapolation to the zero gate width
and thus a determination of the electronic deadtime~\cite{Gaskell:2003dt,aV:2007np}. 
In the present experiment the electronic deadtime was
0.49\% on average.
\par
The computer dead time was calculated from the ratio of
pretriggers $N_{pre}$ to the number of triggers $N_{trig}$ created programmable  module.
Recalling \fig~\ref{F:daq_elec},
one can see that this comparison gives a direct measure of the average
percentage of counts which encountered a busy programmable logic module.
The computer dead time  $\frac{N_{pre}}{N_{trig}}$ was 6.8\% on average for this 
experiment and data was corrected on a run-by-run basis.
\par
The efficiency of tracking in the SOS and HMS drift chambers was defined as the 
probability of finding a valid track for a particle identified as a electron (proton)
for the SOS (HMS).
For the SOS, the rates in the focal plane were 10~kHz or lower 
and the tracking efficiency averaged  about 99.5\%. 
With the HMS at more forward angles, the rates in the HMS focal plane were higher and
ranged from 40~kHz to 400~kHz at the most forward angle. A study~\cite{Horn:2007cebaf} of 
the tracking efficiency 
of the HMS drift chambers found a linear fall-off in the tracking efficiency 
with increasing rates at focal plane which was related to increased 
likelihood of multiple tracks. The HMS 
tracking efficiency was 95.2\% when averaged over all kinematic settings.
\par
Because the proton interacts strongly there is a reasonable probability that
it will interact with the nuclei in either the target housing material or the material
that makes up the HMS detection package.  This means that the HMS trigger will have
an inefficiency and this effect is termed ``proton absorption.''
\par
An estimate of the ``proton absorption'' inefficiency
was made with data from an experiment which ran
just after this experiment.  The
physics governing the proton absorption is the nuclear proton-nucleon
interaction.  The proton-proton
interaction cross section $\sigma_{pp}$ varies from about 47 to 42~mb over the
range of incident proton momentum from 2 to 5~\umom~\cite{Yao:2006px}.
For heavier nuclei the cross section can be
approximated as $\sigma_{pp}A^{0.7}$.
Using the measured cross sections to compute the
proton disappearance one obtains that 95\% of the protons are detected
by the HMS.
\par
The spectrometer configuration
with $\theta_{SOS}=50.01^o$ and $\theta_{HMS}=18.00^o$ was used to
measure the proton absorption.
The SOS and HMS central momenta
were 1.74~\umom \  and 4.34~\umom \
respectively, while the beam energy was 5.25~\uenergy.  The data acquisition
system records single arm events from \emph{both} the SOS and HMS spectrometers
in addition to coincidence events.  Given this, one can compare the electron
arm (SOS) elastic events to the coincidence elastic events in the pure elastic
region of invariant energy, 0.9$\le W \le$1.0~\uenergy.  Requiring that the SOS
in-plane angle be $\pm$ 50 mrad 
ensures the HMS acceptance is large enough to detect all expected protons.
Comparing the elastic yields in each spectrometer shows that the proton absorption effect
causes an inefficiency of approximately 4.0$\pm$1.0\%.  That is, the coincidence
case registered 95$\pm$1\% of the single arm events.  This measurement is
in good agreement with the simple prediction and so will be used as an estimate
for the proton absorption effect.
\par
The experimental corrections
are reported in \tab~\ref{T:corrlist}.
\begin{table}[!hbt]
  \begin{center}
   \begin{tabular}{  c | c }
   \hline
   \hline
   Correction  & Size \\ \hline
 proton absorption &  4.0\%  \\ 
 HMS D.C. efficiency & 5.0\%  \\ 
 SOS D.C. efficiency &  0.5\% \\ 
 electronic deadtime & 0.49\%  \\ 
 computer deadtime & 6.8\%   \\ 
   \hline
   \hline
    \end{tabular}
    \caption{\label{T:corrlist}Efficiency corrections for E01-002 extracted cross sections.  The size reported for
is the size (or average size) of the correction factor applied to the data.  The abbreviation ``D.C.'' stands
for drift chamber.}
   \end{center}
\end{table} 
All corrections, except proton absorption, are calculated on a run-by-run
basis, and are given a nominal 0.1\% uncertainty.  This corresponds
to approximately 10,000 events used to calculate the deadtimes.  An experimental
run in E01-002 usually had at least this many events.  The efficiency of the
\v{C}erenkov detector and electromagnetic calorimeter were measured to be
100\%.  The SOS 3/4 trigger efficiency is assumed to be 100\%.
The HMS 3/4 trigger efficiency is taken to be 100\% and assigned a systematic
error of 1.4\%~\cite{Dalton:2008ff}. 

\section{\label{S:xnextract}Extraction of the Cross Sections}
The acceptance and efficiencies of the detectors must be corrected for in the data
analysis.  This necessitates a detailed understanding of how the acceptance
effect modifies the observed number of counts.
\par
Acceptance effects are dealt with by comparing the experimental
yields (after appropriate cuts) to Monte Carlo simulations.  Dividing the experimental
yields with the Monte Carlo yields will remove any acceptance effects assuming that
the acceptance and the cross sections do not change much over the angular bins and
that the acceptance is properly modeled in the Monte Carlo.
\subsection{\label{S:simu}Acceptance Correction and Normalization}
The Hall C simulation package, SIMC, was 
used for both the signal process and elastic radiative background processes.
This package was
developed by many members of the Hall C collaboration and has been tuned to
the appropriate magnet optics and apertures of the HMS and SOS spectrometers
\cite{jA:2001ab}.  After the appropriate data subtractions and
restrictions were made the $\pi^0$ production process was simulated with a constant
differential cross section in the virtual photon-hadron center of
mass.  The number of counts produced by this Monte Carlo simulation
was then compared to the number of counts in the data distributions.
The measured cross section was extracted assuming
that the ratio of these counts is equal to the ratio of the
differential cross section in a particular kinematic bin.
The chief assumption
made by using this method is that either the cross sections do not
vary much over a kinematic bin or the model cross section and the
measured cross section have the \emph{same} functional dependence
over a bin.  The kinematic binning in the current work is such
that the former condition is likely to hold to high accuracy.
\par
The Monte Carlo simulation was carried out for each
configuration of detector settings.  Typically, several detector
settings contributed differently to each kinematic bin, and these were appropriately
combined to obtain the final cross section for each bin.
\par
The number of counts in a kinematic bin were represented as
the sum of signal and background processes.  Indexing the kinematic
bins by $i$ we have:
\begin{equation}\label{E:allcounts}
N^{\pi(exp)}_i = N^{exp}_i - N^{\gamma}_i - N^{acc}_i - N^{tar}_i. 
\end{equation}
The notation is such that $N^{exp}$ is the number of counts observed in the
experiment and is composed of $N^{\pi(exp)}_i$ which is the number from the signal process,
$N^{\gamma}_i$ which is the number from elastic radiative events, $N^{acc}_i$ which
are the accidental counts, and $N^{tar}_i$ which are events which emerge from the
target container materials.  Only the accidental counts are explicitly subtracted
to compute $N^{\pi(exp)}_i$, since the elastic radiative events are removed by kinematic
restrictions and the events from the target materials were found to be negligible.
The errors are assumed to obey
Gaussian statistics, and $\sqrt{N}$ was taken as the error on
the raw counts after data restrictions.  Bins with less than five events
were not reported.  Ultimately, these errors are
rescaled for any correction factors in the analysis.   
\par
The number of counts in each experimental configuration $j$ contributing
to the kinematic bin $i$, is denoted $N^{exp}_{ij}$.  Normalized to the integrated
luminosity and the efficiency corrections for each setting, one has, for each kinematic
bin:
\begin{equation}\label{E:exp_scale}
\tilde{N}^{exp}_i \equiv \sum_j\tilde{N}^{exp}_{ij} = \sum_j\frac{N^{exp}_{ij}}{\mathcal{L}_j \epsilon_j} 
\end{equation}
where, $\mathcal{L}_j$ is the integrated luminosity for the $j^{th}$ setting.
The factor $\epsilon_j$ is the correction for the efficiency and deadtime for the
$j^{th}$ setting, which is the product of individual efficiency contributions.
Generically, the efficiencies can be expanded
as in \eq~\ref{E:eff}.
\begin{equation}\label{E:eff}
\epsilon = \epsilon^{dc} \times \epsilon^{cdt} \times \epsilon^{edt} \times \epsilon^{abs}
\end{equation}
The labels $dc$, $cdt$, $edt$ and $abs$ denote drift chamber, computer deadtime,
electronic deadtime and proton absorption contributions respectively. 
\par
Taking the ratio of experimental to Monte Carlo $\pi^0$ events
was used to quantify the experimental differential cross section.
\begin{equation}\label{E:extract_ratio}
r_{i}\equiv\frac{\tilde{N}^{\pi(exp)}_{i}}{\tilde{N}^{\pi(mc)}_{i}}
\end{equation}
To the extent that the acceptances are properly modeled we have that
$A_j(\Lambda) \sim A_j^{mc}(\Lambda)$, where $A_j(\Lambda)$
represents the acceptance near a kinematic point $\Lambda$ for the $j^{th}$ spectrometer
configuration.  The above ratio then has a simple
interpretation in terms of the data and model differential cross sections. 
\begin{equation}\label{E:xnextractfinal_expl}
\left(\frac{d\tilde{\sigma}}{dE^{\prime}d\Omega_ed\cm{\Omega}_{\pi}}\right)\Bigg|_{i}=r_i
\left(\frac{d\tilde{\sigma}^{mc}}{dE^{\prime}d\Omega_ed\cm{\Omega}_{\pi}}\right)\Bigg|_{i}
\end{equation}
In \peq~\ref{E:mesprod_succinct} and~\ref{E:flux_define_succinct} the 5-fold
cross section was written in terms of the virtual photon cross section and the photon
flux factor.
The photon flux factor will cancel in the
cross section extraction since it is the same on each side of
\eq~\ref{E:xnextractfinal_expl}.
\begin{equation}\label{E:xnextractfinal_cm}
\left(\frac{d\tilde{\sigma}}{d\cm{\Omega}_{\pi}}\right)\Bigg|_{i}=r_i
\left(\frac{d\tilde{\sigma}^{mc}}{d\cm{\Omega}_{\pi}}\right)\Bigg|_{i}
\end{equation}
The extracted differential cross section $d\tilde{\sigma}/d\cm{\Omega}_{\pi}$
must then be corrected for radiative effects on the pion production process
to produce the final reported cross section $d\sigma^{\gamma^{\ast}}/d\cm{\Omega}_{\pi}$.
\par
Equation~\ref{E:xnextractfinal_cm} embodies the method used to extract
center of mass differential cross sections in this work.  First we selected
a cross section in the center of mass to simulate with, then we constructed the appropriate
ratio from the data analysis after all the appropriate subtractions, after
which the differential cross section (without radiative correction) was extracted.
\par
For the analysis of the pion production data at hand, an initial differential
cross section of a constant 1 $\mu$b/Sr was used in conjunction with the
mentioned procedure.  A binning scheme which gave appropriate counting
statistics in each bin was selected in the
kinematic variables $\{W,\cos{\cm{\theta}},\cm{\phi}\}$. 
The current
experimental statistics suggest the binning schemes
reported in \ptab~\ref{T:data_binlq} and~\ref{T:data_binhq}.
\begin{table}[!hbt]
  \begin{center}
   \begin{tabular}{ c | c | c | c }
   \hline
   \hline
 variable  & $W$ (GeV)  & $\cos{\cm{\theta}}$ & $\cm{\phi}$ (rad) \\ \hline
   range   & 1.092$\le W \le$1.412 & -1.0$\le \cos{\cm{\theta}} \le$ 1.0 &
 $0 \le \cm{\phi} \le 2\pi$  \\ 
   bins   &  8   & 10   & 10 \\ \hline
    \hline
    \end{tabular}
    \caption{\label{T:data_binlq}E01-002 analysis binning for low $Q^2$ data.}
   \end{center}
\end{table} 
\begin{table}[!hbt]
  \begin{center}
   \begin{tabular}{ c | c | c | c }
   \hline
   \hline
 variable  & $W$ (GeV)  & $\cos{\cm{\theta}}$ & $\cm{\phi}$ (rad) \\ \hline
   range   & 1.092$\le W \le$1.412 & -1.0$\le \cos{\cm{\theta}} \le$ 1.0 &
 $0 \le \cm{\phi} \le 2\pi$  \\ 
   bins   &  8   & 6   & 6 \\ \hline
    \hline
    \end{tabular}
    \caption{\label{T:data_binhq}E01-002 analysis binning for high $Q^2$ data.}
   \end{center}
\end{table} 
\subsection{\label{S:radcorr}Radiative Corrections}
Elastic radiative contamination to the data has been treated and subtracted as
a background process in \sect~\ref{S:physback}.
The single pion production mechanism, however, can also be accompanied
with radiation and vertex corrections from the initial or final state charged
particles.  The
treatment of these radiations must be different from the
treatment of elastic radiative events because they directly involve
single pion electroproduction.
The electromagnetic structure of these real photon emissions and vertex corrections
are similar on the leptonic current side but more complicated on the meson 
production side, with the possibility of dependence on many more form factors
than the elastic radiative effects.
\par
The purely single pion production and the
single photon processes are illustrated in \peq~\ref{E:mesrad_equn}
and~\ref{E:mespure_equn}.
\begin{equation}\label{E:mesrad_equn}
e + p \rightarrow e^{\prime} + p^{\prime} + \pi^0 + \gamma
\end{equation}       
\begin{equation}\label{E:mespure_equn}
e + p \rightarrow e^{\prime} + p^{\prime} + \pi^0
\end{equation}       
\par
In addition to the hard photon radiations
there are soft photon radiations.  These actually affect
the experimental results since the missing mass resolution of the experiment has a limit
below which one cannot detect an extra radiated photon.  Thus, all the soft
radiations must be included in a consistent manner to obtain a physically measurable
cross section.  The missing mass constraint allows one to limit the maximum energy of
the radiated photons. 
\par
Here the interest is in correcting the experimentally accessible cross section,
which includes the processes of \peq~\ref{E:mesrad_equn} and~\ref{E:mespure_equn}, 
such that it only represents the pure meson production process.  This means one must
remove the effects of soft photon radiations on the pion production amplitude.
A method for doing this has been developed by Afanasev et al.
\cite{Afanasev:2002ee}.  This calculation is
model dependent and a MAID model~\cite{Drechsel:1998hk} is used in this work for the neutral
pion production portion of the relevant diagrams.
The method of \sref~\cite{Afanasev:2002ee} calculates exactly the contributions from the pure
QED portion of the matrix elements up to uncertainty in the hadronic models.
The hadronic models, however, are included in a modular way such that
better models (perhaps constrained by a first iteration of data analysis)
can be included.  The reference does not calculate radiations due to the
hadronic currents and states that these are smaller by an order of magnitude
and contain considerable theoretical uncertainties.  This situation is
understandable given the fact that the hadronic observables are typically
poorly known in any new region of kinematics, and sparsely known in general.
\par
In \sref~\cite{Afanasev:2002ee} the non-covariant approach
of \sref~\cite{Ent:2001hm} is replaced by a covariant approach which
instead of using the maximum radiated energy, $\Delta E_m$, as a parameter
uses the maximum value of the ``inelasticity,'' $v$, which specifies
the boundaries of missing mass to allow in the calculation.  The
missing mass
must be integrated up to the boundary of the inelasticity parameter~\cite{Bardin:1976qa}.
The inelasticity parameter is defined by:
\begin{equation}\label{E:def_vcut}
v \equiv (k_{\pi} + k_{\gamma})^2 - m_{\pi}^2 ,
\end{equation}
for situations where the pions are undetected experimentally.  The parameter
is such that no radiation corresponds to the situation where $v= 0$.  If
all particles were detected then the procedure would have the value of the
inelasticity unambiguously specified with no need for integration.  It is
clear that the minimum value of $v$ is always zero due to the possibility
of radiating a photon with arbitrarily low energy and the maximum value
should correspond to the experimental data selection.  Since in the present 
work pions are selected via missing mass technique
the method described here for
radiative corrections is especially appropriate.
The correction factor which
must be applied to the measured cross section is defined as $\frac{1}{\delta}$,
with:
\begin{equation}\label{E:delta_define}
\delta(v) \equiv \frac{\sigma^{mes}_m}{\sigma^m}
\end{equation}
In \eq~\ref{E:delta_define}, $\sigma^{mes}_m$ is the measured cross section
including soft radiations and $\sigma^m$ is the pure pion production cross section.
This correction factor must be applied to all the measured data in this
work since the born cross section $\sigma^m$ is the one
to be extracted.  Equation~\ref{E:delta_define} explicitly shows that the correction
factor is a function of the inelasticity parameter though it is implicitly
a function of other kinematic variables like $W$ and $Q^2$.    

\subsection{\label{S:radapp}Application of Radiative Corrections}
The radiative corrections of the type discussed in section~\ref{S:radcorr}
were applied after the raw cross sections are extracted.
Typically, for Hall C studies the radiative corrections are
applied implicitly by including them into the simulation package.  In this
method one is comparing radiated to radiated cross sections and the ratio
of the number of counts is taken to be the same as the ratio of two
non-radiated cross sections.  Current codes which compute the radiative
effects~\cite{Afanasev:2002ee} are too computationally intensive to calculate
the full radiative correction on event-by-event basis, so ``peaking approximations''
are used~\cite{Ent:2001hm}.  For exclusive processes this should not
introduce large systematic errors but here we follow the more direct approach
of extracting the uncorrected cross section $d\tilde{\sigma}/d\cm{\Omega}_{\pi}$ and
correcting it to obtain $d\sigma^{\gamma^{\ast}}/d\cm{\Omega}_{\pi}$.
The code EXCLURAD~\cite{Afanasev:2002ee} was used.
The cross section that was extracted has
a pure pion production part added to a pion production plus soft photon radiation part.
This
center of mass cross section has been introduced in \eq~\ref{E:mesprod_succinct}.
Referring to \eq~\ref{E:delta_define} above, the factor which one must apply
to make $d\tilde{\sigma}$ into the final measured Born-level cross section,
$d\sigma^{\gamma^{\ast}}$ is simply $\frac{1}{\delta}$. 
That is, the EXCLURAD~\cite{Afanasev:2002ee} calculated radiative correction.
\begin{equation}\label{E:extracted_xnradcor}
\frac{d\sigma^{\gamma^{\ast}}}{d\cm{\Omega}_{\pi}}
=\frac{1}{\delta}\frac{d\tilde{\sigma}}{d\cm{\Omega}_{\pi}}
\end{equation}
The
code EXCLURAD must be supplied with a model and we used MAID03~\cite{Drechsel:1998hk} as the standard,
extrapolating the response functions to higher $Q^2$ by a dipole factor.
\par
One might be concerned that this procedure is marred by subtle acceptance effects
in the Monte Carlo simulation.  If one relaxes the constraint that the model
and ``data'' should have the same distributions after iteration, then
this is not a problem.  The acceptance functions which the Monte Carlo creates
should be the same for a given set of detected particles and their
respective momenta.  That is, the acceptance should not depend on what other
particles are created in any given reaction.  Therefore, the only possible
problem which can, and will, arise in this procedure is that processes
with different numbers of undetected particles can have non-zero cross sections
in regions where processes with other undetected particles are kinematically
forbidden.  For example, elastic radiative events have a different phase
space than the pure elastic events.  However, one will never seek to measure
a cross section in a kinematically dis-allowed region so the ratios will
never be extracted in those troublesome regions.  The only constraint, then,
is that the simulated process has the same ``measured'' particles and
is kinematically allowed in every region where one wishes to obtain the
final cross section.
\par
\begin{figure}[!htb]
\begin{center} 
\includegraphics[scale=0.7]{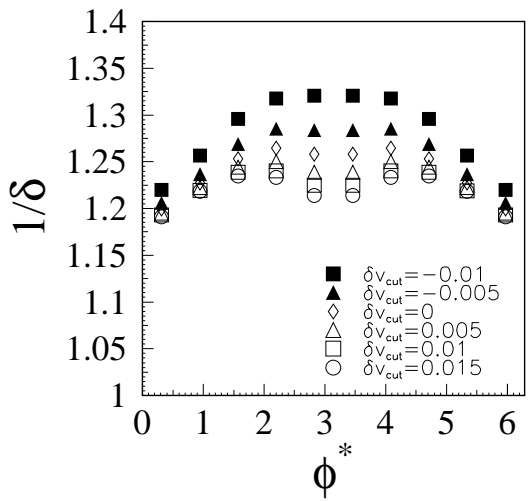}
\caption{\label{F:pirad_1232_cos}Radiative corrections for $W=1.232$~\uenergy,
$Q^2=6.36$~\ufourmomts \ and $\cos{\cm{\theta}}=$ 0.  The inelasticity parameter $v$ was varied to
produce several curves.  The shift of the inelasticity parameter from nominal,
$\delta v_{cut}$ is displayed in the figure.
}
\end{center}
\end{figure}
\begin{figure}[!htb]
\begin{center} 
\includegraphics[scale=0.7]{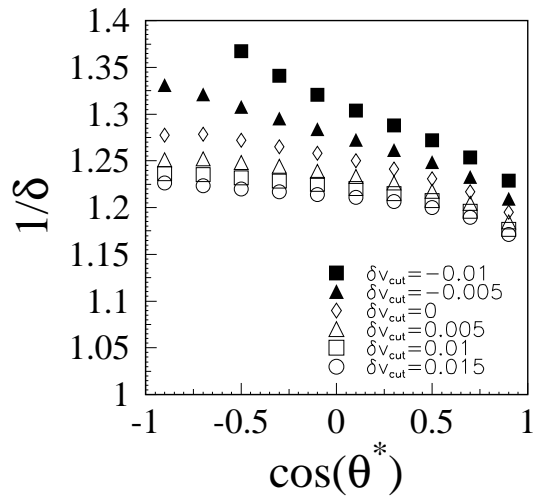}
\caption{\label{F:pirad_1232_phi}Radiative corrections for $W=1.232$~\uenergy,
$Q^2=6.36$~\ufourmomts \ and $\cm{\phi}=\pi$.  The inelasticity parameter $v$ was varied to
produce several curves.  The shift of the inelasticity parameter from nominal,
$\delta v_{cut}$ is displayed in the figure.
}
\end{center}
\end{figure}

Figures~\ref{F:pirad_1232_cos}
and~\ref{F:pirad_1232_phi} display the sizes of the correction factors
in the region of the $\Delta$ resonance as a function of $\cm{\phi}$ for
$\cos{\cm{\theta}}=$ 0 and as a function of $\cos{\cm{\theta}}$ for
$\cm{\phi}=\pi$, respectively.
The parameter $v$ corresponds to the upper bound of the missing mass
restriction shown in \eq~\ref{E:MM_excl}.  Figures~\ref{F:pirad_1232_cos} and~\ref{F:pirad_1232_phi}
show several radiative correction schemes where the parameter $v$ is varied from
the nominal $v_{nom}$.  For the plots we have $\delta v_{cut} \equiv (v-v_{nom})$.
\par
The radiative correction is 20.0-25.0\% in the $\Delta$ region for the
nominal inelasticity values.  Over the entire $W$ range the correction
varies over the somewhat larger range 15.0-27.0\%.  Since a change in the
inelasticity parameter used in the analysis will change the radiative correction, a systematic error should be assigned.  In this case the
corrections for the nominal inelasticity parameters vary very little with
a reasonable sized change in the inelasticity parameter (missing mass
restriction).  The number of data counts, however, is correlated with the
radiative correction through the analysis cuts.  For this reason the error induced
on the final cross section is considered. 
A systematic error of 2.0\% corresponding to the largest deviation
is adopted here.

\section{\label{S:sys}Systematic Errors}
Two types of systematic errors for the measurements were considered.
Some systematics are errors
which affect the cross section data by an overall factor and can be
quantified straightforwardly.  Other systematic
errors are errors which arise from some analysis procedures which
introduce somewhat arbitrary but necessary parameters like the missing
mass acceptance window.  The way the latter type of systematic errors will
be treated is by varying the arbitrary parameters within ``reasonable'' boundaries
and observing the outcomes.
\begin{table}[!hbt]
  \begin{center}
   \begin{tabular}{  c | c }
   \hline
   \hline
   Error  &  Size \\ \hline
 beam current  & 0.5\% \\ 
 proton absorption &   1.0\%  \\ 
 fiducial cuts &   0.5\% \\ 
 collimator cuts  & 0.5\% \\ 
 target boiling &  $<$0.5\%  \\ 
 \v{C}erenkov-calorimeter cut &  1.6\% \\ 
 HMS D.C. efficiency &   0.1\%   \\ 
 SOS D.C. efficiency & 0.1\%  \\ 
 HMS 3/4 trigger efficiency& 1.4\% \\
 electronic deadtime &  0.1\% \\ 
 computer deadtime &  0.1\%  \\ 
 $M_x^2$ cut &  0.35 - 2.8\%  \\ 
 radiative cut &  0.35 - 2.8\%  \\ 
 SOS acceptance &  5.0\%    \\ 
 $\pi^0$ radiative & 1.0-2.0\%   \\ 
 target walls &  1.0\%   \\ 
  \hline
  \hline
    \end{tabular}
    \caption{\label{T:syslist}Systematic errors for the extracted cross sections.
}
   \end{center}
\end{table} 
Table~\ref{T:syslist} displays the systematic errors that were assessed and the
sources which contributed them.  Some of these errors contribute to the
overall normalization of the data and some vary from one data point to
another.  These errors are included as uncertainties on the final
cross section result.  Section~\ref{S:agg_err} quantifies the errors
which vary from point-to-point.
\subsection{\label{S:agg_err}Aggregate Error Estimation}
The point-to-point systematic errors mentioned above require a
sensitivity study because of the fact that the error
does not have a straight forward multiplicative effect
on the cross section data.  The cause of these errors
is aggregate in some sense, built up by the use of several
physically arbitrary (or unknown) but practically necessary parameters.
\par
The three data analysis techniques in this work which produce this
type of systematic error are the $\pi^0$ particle identification,
the elastic radiative rejection and the $\pi^0$ radiative correction.
The $\pi^0$ particle identification uses a missing mass acceptance
width to select the appropriate events, the elastic radiative
rejection uses empirically defined curves to reject background
and the radiative correction uses a radiated energy
parameter, $v$, and a model pion production
cross section.
\par
\begin{figure}[!htb]
\begin{center} 
\includegraphics[scale=0.53]{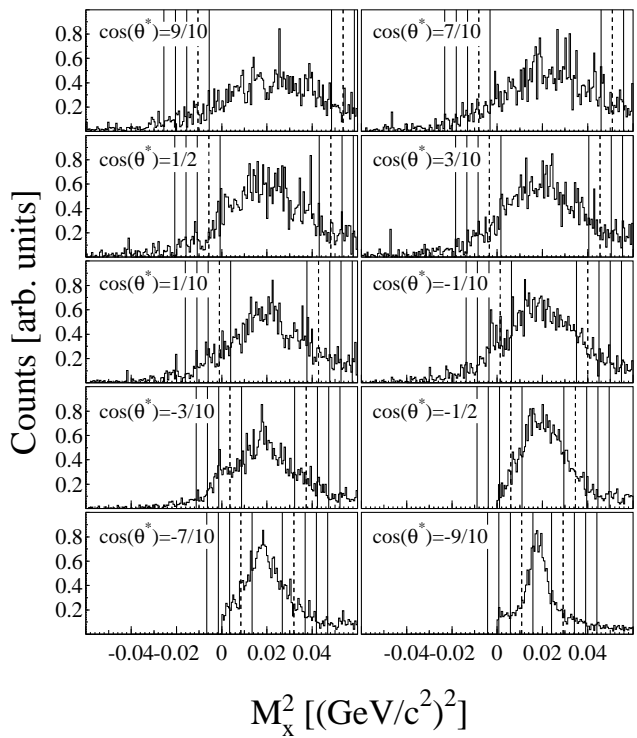}
\caption{\label{F:mm2_syscuts}Various missing mass restrictions used to probe the
systematic error of the nominal width.  The dashed lines represent
the nominal width.
The vertical scale of the plots displays the number of events normalized to the maximum bin content of each histogram.
}
\end{center}
\end{figure}

By using several variations of the missing mass
restrictions one can observe how the cross section
will change.
Figure~\ref{F:mm2_syscuts} shows the various restrictions used to estimate this
error.  
The variation of the cross sections and other extracted observables were
monitored with \emph{only} the missing mass restrictions varied
from the nominal values.  If a simple straight line is fit through the
values one can get a determination of the local first order rate of change
of the quantities of interest.  The scale of the width increments is used
as a multiplier for the approximate error.  For small variations and 
a generic $f(a)$:
\begin{equation}\label{E:sys_taylor}
f(a) \simeq f(a_0) + \frac{df}{da}\bigg|_{a=a_0} \delta_a.
\end{equation}
The symbol $a_0$ represents the nominal parameter and $\delta_a$ the approximate
scale on which $f$ changes, taken to be the parameter increments. The correction
term in \eq~\ref{E:sys_taylor} is then taken to be the systematic error
on the measured cross section due to the missing mass restriction.  Typical
values of this error range from 0.35\% to 2.8\%.
\par
\begin{figure}[!htb]
\begin{center} 
\includegraphics[scale=0.47]{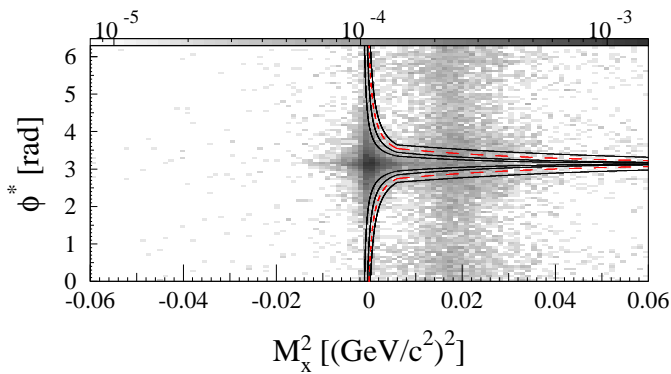}
\caption{\label{F:phimm_syslow}(color online) Radiative rejection overlay for lowest $\cos{\cm{\theta}}$ bin.
Several variations on the nominal rejection curves are displayed.  The dashed
curves are those using the nominal parameters.
} 
\end{center}
\end{figure}
\begin{figure}[!htb]
\begin{center} 
\includegraphics[scale=0.47]{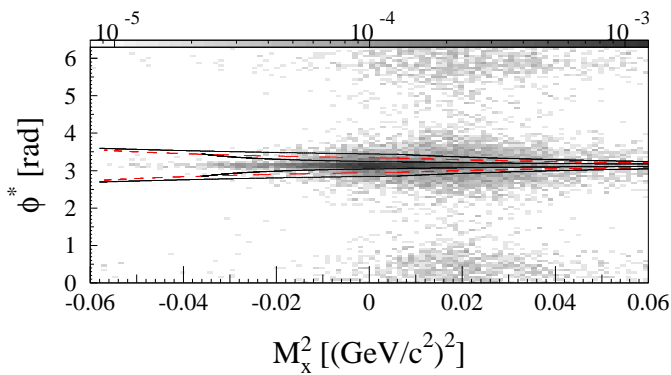}
\caption{\label{F:phimm_sysmed}(color online) Radiative rejection overlay for intermediate and large $\cos{\cm{\theta}}$ bin. 
Several variations on the nominal rejection curves are displayed.  The dashed
curves are those using the nominal parameters.
} 
\end{center}
\end{figure}

An exactly analogous study was carried out for the elastic radiative
rejection.  Recall from \sect~\ref{S:physback} that the radiative
rejection is broken up into three regions of $\cos{\cm{\theta}}$.
Each region has a two dimensional restriction.  Varying these
parameters within reasonable boundaries one can come up with the curves
displayed in \pfig~\ref{F:phimm_syslow} and~\ref{F:phimm_sysmed}.
The errors due to the uncertainty in these rejection cuts are in the range 0.35 - 2.8\%,
which are  very similar to the uncertainties induced by the
missing mass cuts.

\section{\label{S:mult_extract}Extraction of Multipoles}
In Eq.~\ref{E:diffpi_domega_succinct}, the
$\cm{\phi}$ dependence of the differential cross section is explicit, but the  $\cos{\cm{\theta}}$
dependence is not so easily constrained unless one restricts oneself to
states of definite angular momentum or at least states with some finite and
small set of definite angular momentum contributions.  
\par
An empirical fitting procedure is used to extract information about the
P$_{33}$ or $\Delta(1232)$ resonance in the present work.
Multipole amplitudes of the Chew-Goldberger-Low-Nambu (CGLN~\cite{Chew:1957tf}) type: $E_{l\pm}$, $M_{l\pm}$ and $S_{l\pm}$
were extracted where $l$ is the orbital angular momentum of the final state
and the final state nucleon spin is denoted by $\pm$. 
The procedure
hinges on assuming a dominant magnetic dipole, $M_{1+}$, amplitude and assuming
that one has only s- and p-wave contributions to the differential cross section.
\subsection{\label{S:sp_expand}Expansion with s- and p-waves}
The working assumption for the empirical fit is that in the partial wave
series expansion only s- and p-waves will contribute. 
Indeed, the 
$\Delta(1232)$ resonance is a p-wave resonance, P$_{33}$ in
spectroscopic notation.  The next higher excitation, the
P$_{11}(1440)$, or ``Roper'' resonance is also p-wave.  The lowest
lying excitation which decays into a d-wave is the D$_{13}(1520)$.
Furthermore, the underlying non-resonant backgrounds are believed to be s- and p-wave
dominated at these low excitation energies. 
\par
Thus, the $\cos{\cm{\theta}}$ dependent cross sections
can be written explicitly and fit
to experimental data.  The dependence that one obtains by including only the
lowest two final state pion angular momentum contributions is well known
\cite{Drechsel:1992pn,vF:1998ed}.
It is then possible to write the s- and p-wave expansion in terms of three 
unknown functions which depend on $W$ and $Q^2$ and are well defined
functions of $\cos{\cm{\theta}}$ but not functions of $\cm{\phi}$.  
\begin{equation}\label{E:diffpi_cm_expand}
\begin{aligned}
\frac{d\sigma^{\gamma^{\ast}}}{d\cm{\Omega}_{\pi}} =& A(\cos{\cm{\theta}})
 + \epsilon
B(\cos{\cm{\theta}}) \cos{2\cm{\phi}} \\ &+ 
\sqrt{2\epsilon(1+\epsilon)} C(\cos{\cm{\theta}}) \cos{\cm{\phi}}
\end{aligned}
\end{equation} 
The $\sigma_L$ and $\sigma_T$ contributions get combined into one
parameter, $A$, since the present experiment does not vary $\epsilon$ at
a fixed value of $Q^2$ and
therefore cannot separate these contributions.  Using the truncated partial
wave expansion one can then write the explicit angular dependence.
\begin{equation}\label{E:thet_ang_dep}
\begin{aligned}
A(\cos{\cm{\theta}}) &\equiv A_0 + A_1 \cos{\cm{\theta}} + A_2 \cos^2{\cm{\theta}} \\
B(\cos{\cm{\theta}}) &\equiv B_0 \sin^2{\cm{\theta}} \\
C(\cos{\cm{\theta}}) &\equiv (C_0 + C_1\cos{\cm{\theta}})\sin{\cm{\theta}}
\end{aligned}
\end{equation}
\par
\begin{figure*}[!htb]
\begin{center}
\rotatebox{0}{\includegraphics[scale=0.98]{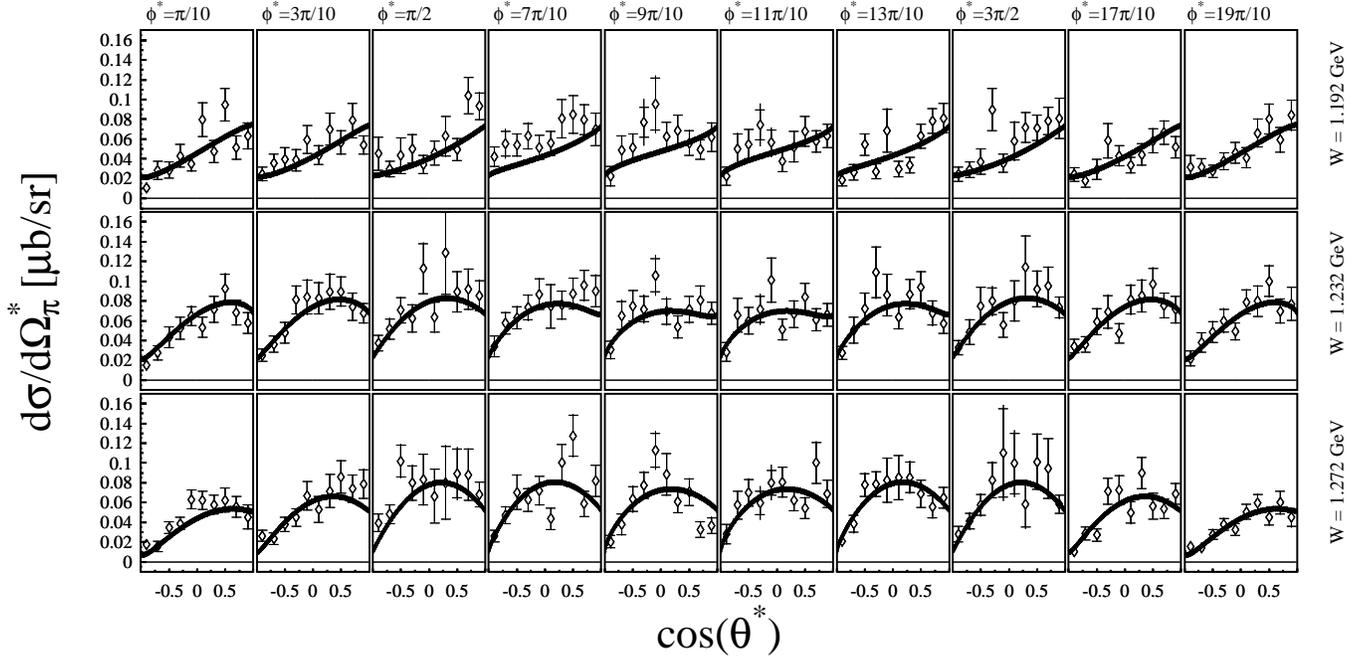}}
\caption{\label{F:sample_81010}Fit to differential cross sections
at $W=1.192, 1.232$ and $1.272$~\uenergy \ with $Q^2=6.43, 6.36$ and $6.29$~\ufourmomts \ respectively.  The
data is binned in $\cm{\phi}$ and displayed as a function of $\cos{\cm{\theta}}$.
Outer error
bars are systematic.
The values of $\chi^2/n_{dof}$ for these fits are 1.31, 0.79 and 1.46
respectively.
}
\end{center}
\end{figure*}

The parameters $A_i$, $B_i$, and $C_i$ are now only functions of the electron
variables $W$ and $Q^2$, and not functions of the hadronic center of mass
angles.  A simple way to proceed in characterizing the extracted cross sections
is to fit the angular distributions in each $W$ and $Q^2$ bin independently.
This point of view is taken in this section, since to include the $W$
and $Q^2$ dependence in the fitting procedures requires detailed knowledge of
the dynamics at least at a level where one can add many resonance and background
contributions with enough free parameters to obtain a physically realistic
parameterization.
Figure~\ref{F:sample_81010} shows an example of fit results using the
$Q^2=$ 6.4~\ufourmomts \ experimental data and the $W=1.192, 1.232$ and $1.272$~\uenergy \ center of mass energy bins.  The
previously described fit is superimposed onto the data. 
\begin{figure*}[!htb]
\begin{center} 
\rotatebox{0}{\includegraphics[scale=0.98]{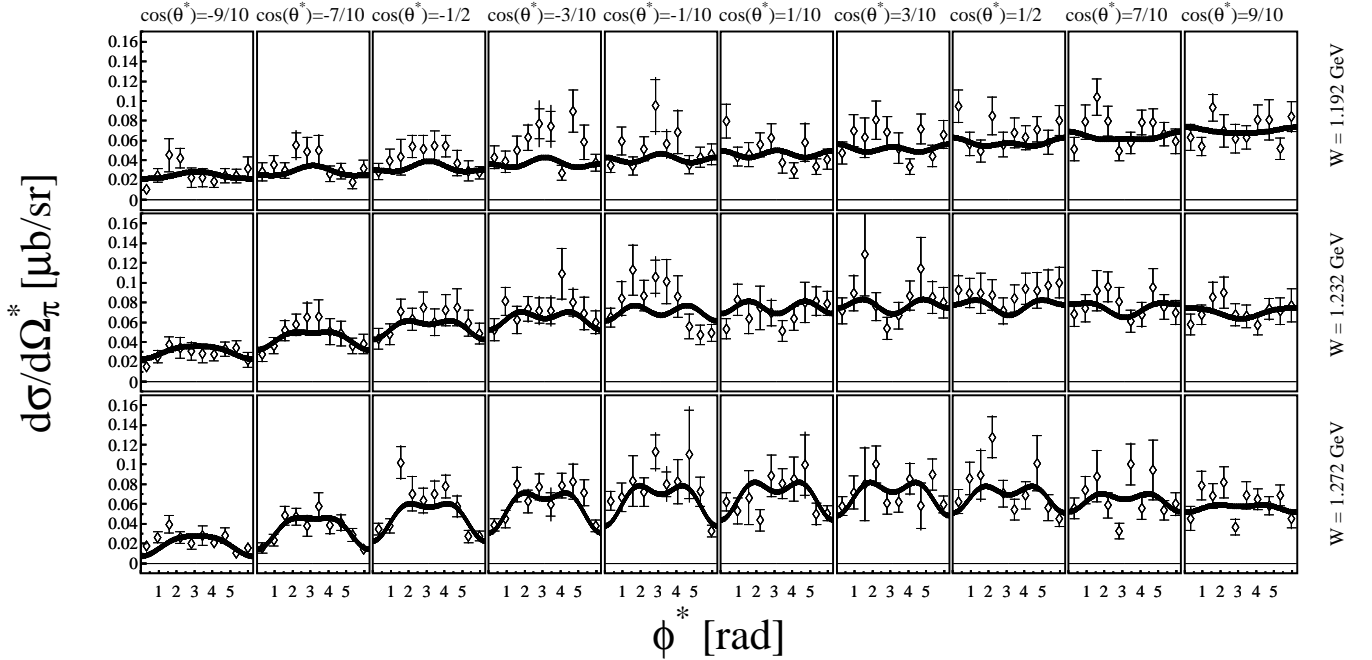}}
\caption{\label{F:sample_81010_phi}Fit to differential cross sections
at $W=1.192, 1.232$ and $1.272$~\uenergy \ with $Q^2=6.43, 6.36$ and $6.29$~\ufourmomts \ respectively.  The data is binned in
$\cos{\cm{\theta}}$ and displayed as a function of $\cm{\phi}$.  Outer
error bars are systematic.
The values of $\chi^2/n_{dof}$ for these fits are 1.31, 0.79 and 1.46
respectively.
}
\end{center}
\end{figure*}
This illustrates the procedure for energy independent analysis of the
differential cross sections.  The binning represented in \fig
\ref{F:sample_81010} shows 40~MeV wide W bins centered on $W=1.192, 1.232$ and $1.272$~\uenergy \ along with
ten angular bins in $\cos{\cm{\theta}}$ and $\cm{\phi}$.
Since
the measurement is unpolarized one should observe only symmetric distributions 
in $\cm{\phi}$.
\par
Figure~\ref{F:sample_81010_phi} displays data points
which are (to within statistical accuracy) symmetric about the point
$\cm{\phi}=\pi$.  This fact is a good check on any extracted cross section.
The $\cm{\phi}$ symmetry is a general feature of the cross section data for
the present experiment.
\par 
All of the other experimental observables can be extracted from these types
of fits by assigning certain physical significance to the fit parameters. 
The extracted cross sections will be made available through Jefferson
Lab for various world data fits or other scientific purposes.  The extracted cross
sections are also displayed in \ptab~\ref{T:xntabl} and~\ref{T:xntabh} in the
Appendix.
\subsection{\label{S:mult_fit}Multipole Fitting}
The fit parameters used in the last section had but one assumption in their
use, namely, that they included only up to p-wave contributions.  The
$\chi^2$ parameters for these fits are fairly good and
therefore one has confidence for at least the low $Q^2$ settings that
s-wave and p-wave contributions approximate the cross section well.
\par
One
can now attempt to go further in the interpretation of these parameters
by constraining the CGLN multipoles.
The $M1$ dominance procedure~\cite{dhL:1978ds,vF:1998ed} has traditionally
been employed to reduce the number of contributing multipoles in the s- and p-wave
amplitudes so that they can be extracted from fits to the angular distributions.
If one assumes that the $M_{1+}$ multipole dominates, the s- and p-wave fit
parameters can be related to the multipole ratios in a simple way.
\begin{widetext}
\begin{equation}\label{E:m1dominance}
\begin{aligned}
A_0&=\cgln|M_{1+}|^2\left[\frac{5}{2}-3\frac{\re{E_{1+}^{\ast}M_{1+}}}{
|M_{1+}|^2}+\frac{\re{M_{1+}^{\ast}M_{1-}}}{|M_{1+}|^2}\right] \\
A_1&=\cgln|M_{1+}|^2 2\frac{\re{E_{0+}^{\ast}M_{1+}}}{|M_{1+}|^2} \\
A_2&=\cgln|M_{1+}|^2 \left[-\frac{3}{2}+9\frac{\re{E_{1+}^{\ast}
M_{1+}}}{|M_{1+}|^2}-3\frac{\re{M_{1-}^{\ast}M_{1+}}}{|M_{1+}|^2}\right]\\
B_0&=\cgln|M_{1+}|^2 \left[-\frac{3}{2}-3\frac{\re{E_{1+}^{\ast}
M_{1+}}}{|M_{1+}|^2}-3\frac{\re{M_{1-}^{\ast}M_{1+}}}{|M_{1+}|^2}\right]\\
C_0&=\cgln|M_{1+}|^2\qfac\frac{\re{S_{0+}^{\ast}M_{1+}}}{|M_{1+}|^2} \\
C_1&=\cgln|M_{1+}|^2 6\qfac\frac{\re{S_{1+}^{\ast}M_{1+}}}{|M_{1+}|^2} \\
\end{aligned}
\end{equation}
\end{widetext}
In \eq~\ref{E:m1dominance}, $\cm{\mathbf{k}}_{\pi}$ is the pion momentum in
the center of mass,
$\cm{\mathbf{q}}$ is the virtual photon momentum in the center of mass and
$m_p$ is the proton rest mass. 
There are six combinations of multipoles, all involving $M_{1+}$.  By
substitution for the six parameters in \eq~\ref{E:diffpi_cm_expand} the
differential cross section can be expressed in terms of the leading $|M_{1+}|^2$
and the five interference terms in \eq~\ref{E:m1dominance}.  Then, the
experimental differential cross sections can be fit to extract these six multipole
combinations.
\par
\begin{figure}[!htb]
\begin{center} 
\includegraphics[scale=0.45]{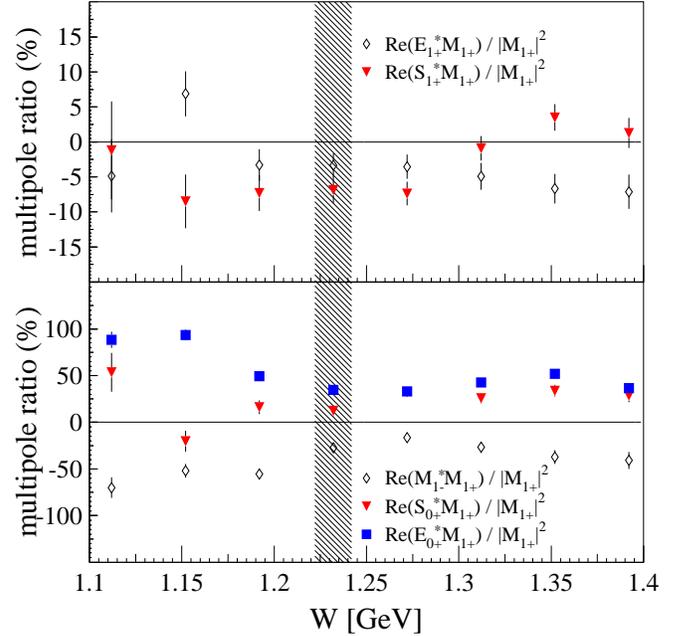}
\caption{\label{F:allmult_lowq}(color online) Results of the simple $M_{1+}$ dominance fit for
the nominal $Q^2=$ 6.4~\ufourmomts \ data set as a function of invariant energy $W$.  The region
of the $\Delta(1232)$ resonance is shaded.
}
\end{center}
\end{figure}
\begin{figure}[!htb]
\begin{center} 
\includegraphics[scale=0.45]{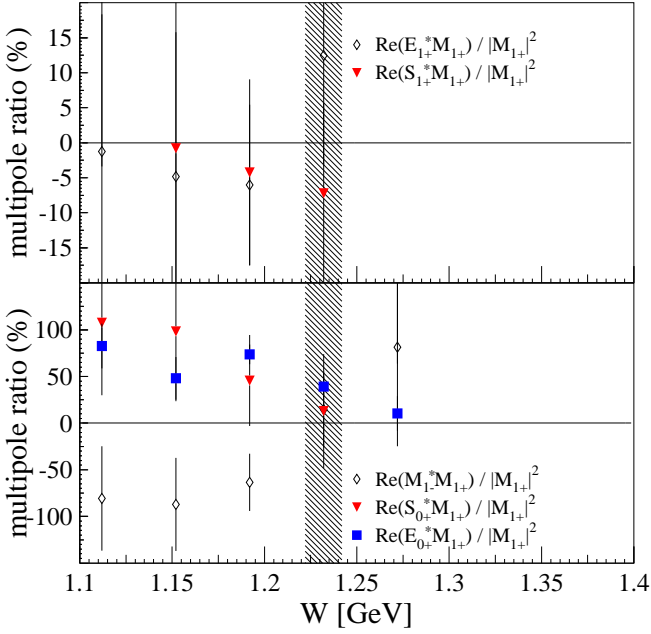}
\caption{\label{F:allmult_highq}(color online) Results of the simple $M_{1+}$ dominance fit for the
nominal $Q^2=$ 7.7~\ufourmomts \ data set as a function of invariant energy $W$.  The region of
the $\Delta(1232)$ resonance is shaded.
}
\end{center}
\end{figure}

The results
of the $W$-independent fits for the current experimental data
sets are shown in \pfig~\ref{F:allmult_lowq} and~\ref{F:allmult_highq}.
\par
From \fig~\ref{F:allmult_lowq} it is seen that even near the
$\Delta(1232)$, unlike the situation for low $Q^2$, the assumption
of $M_{1+}$ dominance is only approximate.  This is due to a
combination of factors.  The $\Delta$ resonance is known to fall off
with $Q^2$ more steeply than other resonances and background.  The
amplitudes of the Roper (P$_{11}(1440)$) resonance, while very small at
low $Q^2$ have recently been shown to become large~\cite{Aznauryan:2004jd} with increasing $Q^2$.
These phenomena are manifested in the relative importance
of $\re{M_{1-}^{\ast}M_{1+}}/|M_{1+}|^2$
and $\re{E_{0+}^{\ast}M_{1+}}/|M_{1+}|^2$ in \fig~\ref{F:allmult_lowq}.
\par
The values of $R_{EM}$ and $R_{SM}$ extracted at the $W=1.232$~\uenergy \ and
$Q^2=6.4$~\ufourmomts \ are
listed in \tab~\ref{T:results}.  This includes systematic errors in the extracted multipole parameters.  These
are extracted with similar methods to those presented in \sect~\ref{S:sys_multext}.
The values of $R_{EM}$, modulo the
caveats given above, are somewhat consistent with previous values which are negative
and small in magnitude.  The value of $R_{SM}$ is more controversial.  Previous
analyses of the world data by two different schemes, JINR~\cite{Aznauryan:2004jd} and
MAID~\cite{Drechsel:1998hk} yield differences of a factor of two. 
\par
As seen in \fig~\ref{F:allmult_highq} the fits for multipole amplitudes
at the higher $Q^2$ ($W=1.232$~\uenergy, $Q^2=7.7$~\ufourmomts) are poorly
constrained, since this data has less statistical significance and poor
angular coverage.  One can only say that it is likely that $R_{EM}$ continues to
be small.
\par
The angular integrals, $\sigma$, of the differential cross sections $d\sigma^{\gamma^{\ast}}/d\cm{\Omega}_{\pi}$
were calculated
in terms of the fit parameters.  The errors on the
fit parameters can then be propagated through to this integrated cross
section.  This method is the most consistent way of displaying
the desired cross sections with the detector acceptance
effects removed.  The method is subject to large uncertainty
when the data points have incomplete angular distribution.  For the
present data set this happens for the high $W$
points at higher $Q^2$. 
\par
\begin{figure}[!htb]
\begin{center} 
\includegraphics[scale=0.7]{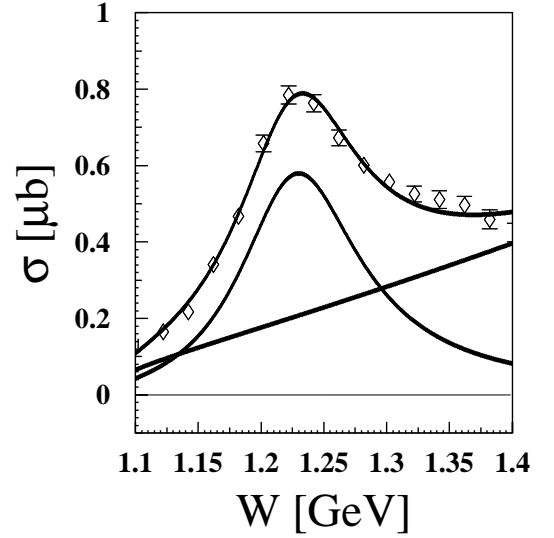}
\caption{\label{F:wfit_lowq}The total angle-integrated cross section with 
$Q^2\sim$ 6.4~\ufourmomts \ at the $\Delta$ peak. Cross sections are fit with a Breit-Wigner function and a fractional
power polynomial of first order.
}
\end{center}
\end{figure}
\begin{figure}[!htb]
\begin{center} 
\includegraphics[scale=0.7]{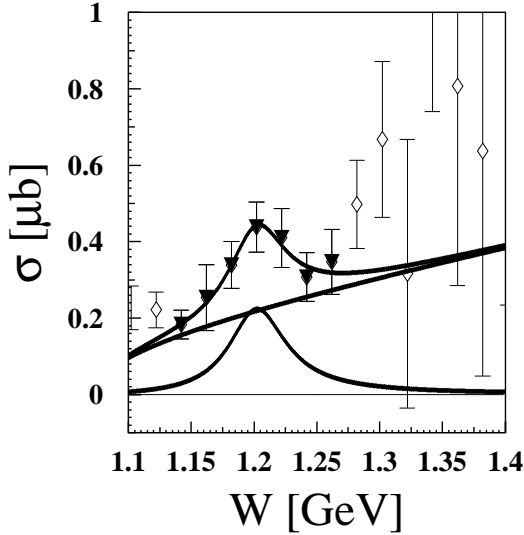}
\caption{\label{F:wfit_highq}The total angle-integrated cross section with 
$Q^2\sim$ 7.7~\ufourmomts \ at the $\Delta$ peak. Cross sections are fit with a Breit-Wigner function and a fractional
power polynomial 
of first order.  The dark inverted triangle data
points are the only ones used to constrain the fit due to lack of statistics
for higher $W$ and the possibility of small elastic radiative contamination
at lower $W$.
}
\end{center}
\end{figure}

Figures~\ref{F:wfit_lowq} and~\ref{F:wfit_highq} display the experimentally
observed angle-integrated cross sections and fits to them.  For these
figures the differential cross section with 16 $W$ bins and 49 angular
bins were fit using the previous parameters.  The angular bins included
7 $\cm{\phi}$ bins and 7 $\cos{\cm{\theta}}$ bins.
The fits to the $W$ behavior include
a resonance contribution with the appropriate threshold behavior~\cite{Stoler:1993yk} and a
polynomial background of various order.  The specific function used to fit the resonance
and background was the following:
\begin{equation}\label{E:res_form}
\begin{aligned}
\sigma &= \frac{fa_0^2}{(W^2-a_1^2)^2+a_2^2} + p(W-W_{th},\{a_n\}_3^N) \\
f&\equiv \sqrt{\left(\frac{W^2+m_{\pi}^2-m_{p}^2}{2W}\right)^2 - m_{\pi}^2}.  
\end{aligned}
\end{equation}
The $a_i$ are adjustable parameters and the function $p(W-W_{th},\{a_i\}_3^N)$ represents a polynomial
in $W$ with $N-3$ terms.  To obtain the best fit a polynomial including all non-zero integer and
half-integer powers up to $(W-W_{th})$ was included in \fig~\ref{F:wfit_lowq}.
For \fig~\ref{F:wfit_highq} the same polynomial was used.
One can see that
the background contribution is roughly 50\% and 100\% of the peak height for the lower
$Q^2$ and higher $Q^2$ data respectively.  One should be aware that this rough
determination of the background has large systematic error due to the arbitrary
selection of the type of polynomial to use. 
These factors can be used
as a rough correction factor to the cross section for extraction of $(G_M^{\ast})^2$.  The seemingly
large background contribution for the higher $Q^2$ data indicates
that the resonance may not be dominating the cross section at these
high values of momentum transfer.  The fit is a very rough approximation
and detailed procedures with more physical inspiration
are discussed at the end of this section.
\par
We extracted the transition form factor $G_M^{\ast}$ from the angle integrated cross
sections evaluated at the $\Delta$ pole position.
The notation which we adopt is that of Jones and Scadron~\cite{Jones:1972ky}
which is based on relativistic current structures in analogy with elastic scattering.
\par
The magnetic form factor $G_M^{\ast}$ will be extractable and
directly related to the multipole $M_{1+}$ if the resonance is completely
dominant at the peak position.  First note that when one integrates
the angular distribution quoted in \eq~\ref{E:diffpi_cm_expand} one obtains
$\frac{2}{3}(A_2+3A_0)$.  This expression can easily be put in terms of the
multipole amplitudes (assuming, still, $M_{1+}$ dominance) to get:
\begin{equation}\label{E:wintxn_mult_relate}
\sigma = 8\pi\cgln |M_{1+}|^2
\end{equation}
One can therefore extract the (presumed dominant) $M_{1+}$ multipole from just
a measure of the angle integrated cross section.  Further, one has the
relation:
\begin{equation*}
\im{M_{1+}^{(3/2)}} = \eta_b \sqrt{\frac{2}{3}}G_{M}^{\ast}.
\end{equation*}
The factor $\eta_b$ serves to relate the magnetic transition form factor
to the multipole amplitude as in \sref~\cite{Burkert:2004sk}.
Therefore a measurement of the
cross section, armed with the assumption that $M_{1+}\simeq i\im{M_{1+}^{(3/2)}}$,
is a direct measurement of the magnetic transition form
factor assuming resonance and magnetic multipole dominance.
\begin{figure}[!htb]
\begin{center} 
\includegraphics[scale=0.7]{figures/new_gm_world.epsi}
\caption{\label{F:current_GM}(color online) Current experimental situation for $G_M^{\ast}$ (in the Jones and Scadron convention) with increasing $Q^2$.  The points are from
CLAS~\cite{Joo:2001tw,Aznauryan:2004jd,Ungaro:2006df},
and Hall C~\cite{Frolov:1998pw}.  The error bars for the $M1$ dominance results are the sum (in quadrature) of
the statistical and systematic errors.  The result from the Aznauryan fit to the
present cross section is also shown~\cite{Aznauryan:2003zg} with
only statistical error bars displayed.
}
\end{center}
\end{figure}
Figure~\ref{F:current_GM} shows the current experimental situation for
the transition form factor including the values extracted in this work.
The
values for $G_M^{\ast}$ are computed by taking the total center of mass
cross section at the $\Delta$ pole position ($W=$ 1.232~\uenergy) from \pfig~\ref{F:allmult_lowq}
and~\ref{F:allmult_highq} and correcting the resonance value using the
fits displayed in \pfig~\ref{F:wfit_lowq} and~\ref{F:wfit_highq}.  According to
these fits the non-resonant background accounts for 50~\% and 100~\% of the resonant
contribution for the $Q^2=$ 6.4  and 7.7~\ufourmomts \ points respectively.
Note that these values are statistically quite well constrained by the data even at the
higher $Q^2$ point.  The results for $G_M^{\ast}$ are compared to the dipole
form factor
$G_D=3(1+Q^2/0.7)^{-2}$ in \fig~\ref{F:current_GM}.
\par
Clearly, although the transition form factor is well constrained statistically,
the magnetic multipole dominance is a rather crude approximation since the
total cross section is not simply due to the $M_{1+}$.  The problem is that
we have neglected all other amplitudes which do not interfere with
$M_{1+}$, such as $|E_{0+}|^2$ and $|M_{1-}|^2$ which certainly contribute significantly to the
total cross section, and should be included in the fit, especially at
these higher values of $Q^2$.
The numerical results of the fit for $G_M^{\ast}$ are given in \tab~\ref{T:results}. The systematic uncertainties in 
$G_M^{\ast}$ introduced by this method have been quantified by obtaining $G_M^{\ast}$ for various assumptions of the background shape, and are significantly greater that the statistical uncertainties.
\begin{table*}[!hbt]
\begin{center}
\begin{tabular}{c|c|c|c}
\hline
\hline
$Q^2$	  &   $G_M^{\ast}/3G_D$		            &	$R_{EM}$(\%)		                    &	$R_{SM}$(\%) \\
\hline
6.36    &  0.307  $\pm$ 0.0033	$\pm$0.058  &	-3.349	$\pm$1.711	$\pm$0.028 &	-6.894	$\pm$1.876	$\pm$0.084\\
7.69    &  0.238  $\pm$0.014	$\pm$0.059  &	12.482	$\pm$15.738	$\pm$0.056 &	-7.217	$\pm$12.819	$\pm$0.020\\
\hline
\end{tabular}
\caption{\label{T:results}The results of the fits to the data as described above. The first error is statistical, the second systematic.}
\end{center}
\end{table*}
\par
A more realistic fitting procedure has been undertaken at Jefferson Lab.  The
analysis uses a unitary isobar model including appropriate non-resonant
background contributions.  The standard isobar approach of Drechsel et al.~\cite{Drechsel:1998hk} 
is complimented by the approach of Aznauryan~\cite{Aznauryan:2003zg}.  The non-resonant background 
consists of Born and t-channel $\rho$ and $\omega$ contributions.  
The most up-to-date world data on nucleon-pion form factors and
higher energy resonance contributions is used as well.  The procedure is
the same as was used to extract the $\Delta$ excitation parameters in \sref~\cite{Ungaro:2006df}.
Table~\ref{T:Azresults} displays the relevant parameters and the errors.
\begin{table*}[!hbt]
\begin{center}
\begin{tabular}{c|c|c|c}
\hline
\hline
$Q^2$	  &   $G_M^{\ast}/3G_D$		            &	$R_{EM}$(\%)		                    &	$R_{SM}$(\%) \\
\hline
6.36    &  0.477  $\pm$ 0.009	$\pm$0.043  &	-1.7	$\pm$1.9	$\pm$1.6 &	-22.3	$\pm$4.4	$\pm$3.4\\
7.69    &  0.404  $\pm$0.024	$\pm$0.056  &	-  &	-\\
\hline
\end{tabular}
\caption{\label{T:Azresults}The results of the Jefferson Lab fitting procedure carried out
by Aznauryan.  The first error is statistical, the second systematic.}
\end{center}
\end{table*}
\subsection{\label{S:sys_multext}Systematic Errors in Extracted Amplitudes}
\begin{figure}[!htb]
\begin{center} 
\includegraphics[scale=0.7]{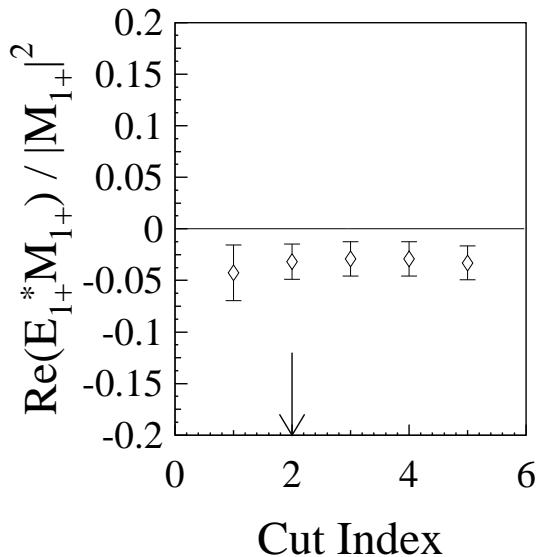}
\caption{\label{F:rem_sys_lowq}The variation of the $R_{EM}$ estimator on missing mass
restriction.  From left to right the restrictions are labeled from
narrowest to widest with the arrow marking the nominal.  The widths
which correspond to this plot
are displayed in \fig~\ref{F:mm2_syscuts}.
}
\end{center}
\end{figure}

One may also be interested
in the systematic error on an observable extracted by a fitting method.
Figure~\ref{F:rem_sys_lowq} shows how our estimator for $R_{EM}$ varies
due to the missing mass restriction.
An estimate for the uncertainty on $R_{EM}$ due to the missing mass restriction
is 1.0\%, based on this analysis.  The systematic errors on the other extracted
multipoles are evaluated in the same way.

\section{\label{S:comp_world}Results in the Context of Previous World Data}
Contributions to the previous world data which are noted here are the following.
At lower $Q^2$ data have been obtained by the MAMI (Mainzer Microtron)
\cite{Pospischil:2000ad,Ahrens:2004pf,Elsner:2005cz,Stave:2006ea,Sparveris:2006uk}, ELSA
(University of Bonn)
\cite{Bantes:2003ta}, LEGS (Brookhaven) collaboration
\cite{Sandorfi:1998xr,Blanpied:1997nd} and the BATES (MIT)
collaboration
\cite{Sparveris:2004jn}.
The Jefferson Laboratory spectrometer
Hall A~\cite{Kelly:2005jj}
has also made a significant
contribution to the question of the structure of the $N\rightarrow\Delta$
transition.  CLAS (Jefferson Laboratory) has also obtained a  large
amount of data over a large range in $Q^2$
\cite{Joo:2001tw,Aznauryan:2004jd,Ungaro:2006df}.  Jefferson Laboratory
Hall C~\cite{Frolov:1998pw}, at $Q^2=2.8$ and $4.0$~\ufourmomts \ was the
predecessor to the present experiment.
\subsection{\label{S:rem}The Electric Quadrupole to Magnetic Dipole ratio $R_{EM}$}
\begin{figure}[!htb]
\begin{center} 
\includegraphics[scale=0.7]{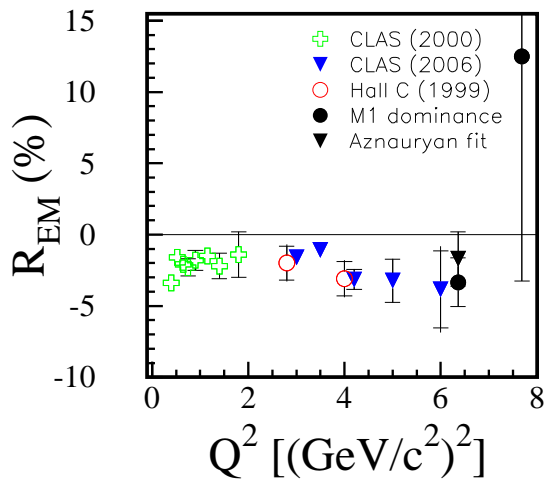}
\caption{\label{F:previous_REM}(color online) Current experimental situation for $R_{EM}$ with increasing $Q^2$.  The points are from
CLAS~\cite{Joo:2001tw,Aznauryan:2004jd,Ungaro:2006df},
and Hall C~\cite{Frolov:1998pw}.  The outer error bars for the $M1$ dominance results are the sum (in quadrature) of
the statistical and systematic errors.  The result from the Aznauryan fit to the
present cross section is also shown~\cite{Aznauryan:2003zg} with
only statistical error bars displayed.
}
\end{center}
\end{figure}

Figure~\ref{F:previous_REM} shows the status of the world data on $R_{EM}$, including
the present result obtained by the M1 dominance method and the more
sophisticated Jefferson Lab (Aznauryan) fit.  The real photon point
at $Q^2=0$ is small in magnitude and negative in sign.  This situation shows
no drastic change up to $Q^2$ of about 7.7~\ufourmomts.
\par 
The results for $R_{EM}$ indicate that the baryon helicity non-conserving
element is on the order of two times as great as the baryon helicity conserving
element.  Perturbative QCD predicts that the helicity non-conserving element vanishes,
causing $R_{EM} \rightarrow 1$~\cite{Brodsky:1981kj,Carlson:1985mm}. 
In the realm of our simplified multipole extraction procedure and also that of
a unitary isobar fit,
one therefore finds that the data indicates the pQCD limit has not yet been
reached for $\Delta$ excitation.  
\subsection{\label{S:gm}Magnetic Form Factor $G_M^{\ast}$}
Figure~\ref{F:current_GM} shows the status of the world data on $G_M^{\ast}$
relative to the dipole form factor $G_D=3(1+Q^2/0.7)^{-2}$.  For
the present data the result is obtained by the methods discussed in
\sect~\ref{S:mult_fit}.  At lower $Q^2$ the resonance is quite
strong and the $M_{1+}$ multipole dominates neutral pion
production in the vicinity of the $\Delta$ resonance pole, so that
$G_M^{\ast}$ form factors which have been extracted by a variety of
approaches yield rather similar results.  However, at high $Q^2$ the rapid
decay of $G_M^{\ast}$ relative to the non-resonant background, and relative
to the increased importance of the tails of other resonances, such as the
P$_{11}(1440)$ (Roper) resonance requires one to make a careful analysis in the
framework of all the available data.  This has been the goal of several analysis
groups including MAID, SAID and JINR.
\par
\begin{figure}[!htb]
\begin{center} 
\includegraphics[scale=0.7]{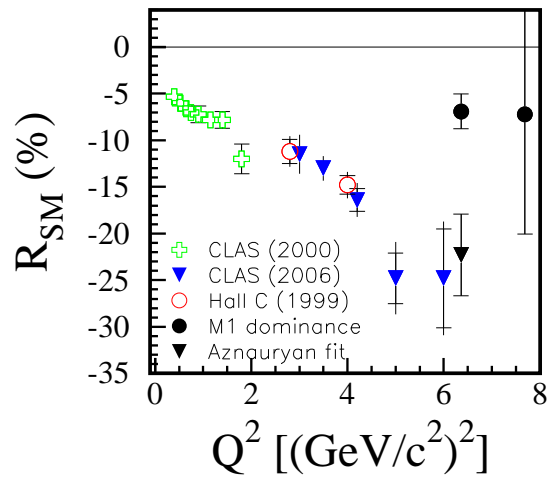}
\caption{\label{F:previous_RSM}(color online) Current experimental situation for $R_{SM}$ with increasing $Q^2$.  The points are from
CLAS~\cite{Joo:2001tw,Aznauryan:2004jd,Ungaro:2006df},
and Hall C~\cite{Frolov:1998pw}.  The outer error bars for the $M1$ dominance results are the sum (in quadrature) of
the statistical and systematic errors.  The result from the Aznauryan fit to the
present cross section is also shown~\cite{Aznauryan:2003zg} with
only statistical error bars displayed.
}
\end{center}
\end{figure}

Overall, $G_M^{\ast}$ is falling much faster than the dipole form factor $G_D$ in the
previously measured $Q^2$ region from 0 to 5 or 6~\ufourmomts.  Asymptotically, if
the pQCD constituent scaling rules were operative, this form factor should begin
to scale as $1/Q^4$, as the dipole form factor does, so the constituent
scaling does not yet occur for $G_M^{\ast}$.  This is consistent with the result for
$R_{EM}$.  The helicity non-conserving amplitude should dominate $G_M^{\ast}$
whenever $R_{EM}$ is small.  Thus, the data on $G_{M}^{\ast}$ and $R_{EM}$ are consistent.
One may then speculate that when $R_{EM}$ becomes large and positive, the $G_M^{\ast}$ may
begin to scale. 
For completeness, the current experimental situation for $R_{SM}$ is shown in \fig~\ref{F:previous_RSM}.
It seems that the M1 dominance extraction procedure used here is especially questionable
in the case of $R_{SM}$ but the Jefferson Lab procedure yeilds results which are consistent
with previously extracted values at lower $Q^2$.

\section{\label{S:concl}Conclusion}
This work has accomplished several goals.  The first and foremost goal
was to extract the center of mass neutral pion electroproduction cross
section in the invariant mass region roughly corresponding to the
well-known $\Delta$ resonance.  This goal was accomplished and the systematic
errors on the cross section were evaluated using the best current knowledge
of the detector systems and analysis procedures.  The next goal of the
analysis was to investigate (in a simplified way) what the cross sections
suggest for the most important multipole amplitudes and transition
form factors relating to the measured process and in particular to the
$\Delta$ resonance.
\par
In the realm of a fit which includes only s-wave and p-wave contributions
and assumes that the multipole $M_{1+}$ dominates all other multipoles
(an assumption which seems to be challenged by the size of $E_{0+}$),
with the $\Delta$ resonance being dominant at the resonance position, one
can obtain values for $R_{EM}$, $R_{SM}$ and $G_M^{\ast}$.  The specifications
of $R_{EM}$ and $R_{SM}$ depend in detail on the angular distribution of the
cross section and thus are only well determined for the $Q^2=6.4$~\ufourmomts \ 
data set.  The most significant facts that are obtained using these
methods are that  $R_{EM}=-3.3\pm1.7$\% and that $G_M^{\ast}$ seems to be
still dropping much faster than the simple dipole form, suggesting that there are
soft mechanisms in the $\Delta$ excitation which are still important
\cite{Carlson:1998bg}.
\par
It is important to reiterate the physical effects which were not considered
in this work.  Everything assumes that single photon exchange is completely
appropriate for dealing with observables measured to the accuracy that
was displayed in this data.  This is probably a good assumption but there
are some places, including calculation of radiative effects which two
photon exchange mechanisms might have a more important role.  These
are left to subsequent study.  There has been some recent work on the
subject and it is hopeful that the two photon effects can be understood in detail
and corrected for in the future if necessary~\cite{Carlson:2006tp}.  In extracting our estimates for the
multipole ratios and transition form factor of the $\Delta$ excitations
it was assumed that the
$\Delta$ dominates at the resonance position, though this assumption has already
been shown to be suspect in \sect~\ref{S:mult_fit}.  Contributions from
multipoles other than $M_{1+}$, specifically $E_{0+}$ may also be showing their
importance and thus spoiling the M1 dominance that was assumed. 
\par
The more physically motivated fitting procedure explored produced results for
$G_M^{\ast}$ and $R_{EM}$ which were consistent with the interpretations from the simpler
M1 dominance procedure.  Specific values for these parameters are modified
in this procedure and, in particular, $R_{EM}=-1.7\pm1.9$\% for $Q^2=6.4$~\ufourmomts. 
$R_{SM}$ becomes much smaller than the M1 dominance result and this fact is not
reproduced in other unitary isobar fitting procedures such as MAID currently. 
\par
These conclusions seem to indicate that the studied process is not in
a regime where perturbative QCD is dominating behaviour.  Continued investigation
is necessary  
to help uncover the inner workings of
hadronic physics and QCD, especially in this intermediate energy region between
hadronic and quark descriptions.
\par
As for measurements at higher $Q^2$, this will have to await the
completion of the Jefferson Lab 12~\uenergy~upgrade.

\appendix*
\section{Tables of Differential Cross Sections}
The measured differential cross sections mentioned throughout the text are listed here.  The author
can be contacted for an electronic version of these data points.  
\newpage
\begin{widetext}
\begin{center}
\begin{longtable}{ccccccccc}
  \caption{\label{T:xntabl}Differential cross sections for the small SOS spectrometer angle.} \\
  \hline
  \hline
$Q^2$ &$\epsilon$ &$W$ &$\cos{\cm{\theta}}$ &$\cm{\phi}$ &$d\sigma^{\gamma^{\ast}}$ &$\sigma_{stat}$ &$\sigma_{norm}$ &$\sigma_{sys}$ \\
(GeV$^2$) &none &(GeV) &none &(rad) &(nb/Sr) &(nb/Sr) &(nb/Sr) &(nb/Sr) \\
  \hline \\[0.05cm]
  \endfirsthead
  \hline
  \hline
$Q^2$ &$\epsilon$ &$W$ &$\cos{\cm{\theta}}$ &$\cm{\phi}$ &$d\sigma^{\gamma^{\ast}}$ &$\sigma_{stat}$ &$\sigma_{norm}$ &$\sigma_{sys}$ \\
(GeV$^2$) &none &(GeV) &none &(rad) &(nb/Sr) &(nb/Sr) &(nb/Sr) &(nb/Sr) \\
  \hline \\[0.05cm]
  \endhead
6.564 &0.4523 &1.112 &-0.9 &-2.827 &5.1456 &3.8602 &0.29671 &0.0022301 \\[0.05cm]
6.564 &0.4523 &1.112 &-0.9 &-1.571 &7.3159 &5.1838 &0.42185 &0.00092151 \\[0.05cm]
6.564 &0.4523 &1.112 &-0.9 &-0.9425 &4.721 &3.3702 &0.27223 &0.0081633 \\[0.05cm]
6.564 &0.4523 &1.112 &-0.9 &-0.3142 &6.7928 &6.7933 &0.39169 &0.0002199 \\[0.05cm]
6.564 &0.4523 &1.112 &-0.9 &0.3142 &4.2895 &4.2898 &0.24734 &0.0023096 \\[0.05cm]
6.564 &0.4523 &1.112 &-0.9 &0.9425 &2.2098 &2.2099 &0.12742 &0.0043183 \\[0.05cm]
6.564 &0.4523 &1.112 &-0.9 &1.571 &2.2215 &2.2216 &0.1281 &0.0026498 \\[0.05cm]
6.564 &0.4523 &1.112 &-0.9 &2.199 &2.6294 &2.6295 &0.15162 &0.0038613 \\
\end{longtable}
 \end{center}
\begin{center}
 \begin{longtable}{ccccccccc}
  \caption{\label{T:xntabh}Differential cross sections for the large SOS spectrometer angle.} \\
  \hline
  \hline
$Q^2$ &$\epsilon$ &$W$ &$\cos{\cm{\theta}}$ &$\cm{\phi}$ &$d\sigma^{\gamma^{\ast}}$ &$\sigma_{stat}$ &$\sigma_{norm}$ &$\sigma_{sys}$ \\
(GeV$^2$) &none &(GeV) &none &(rad) &(nb/Sr) &(nb/Sr) &(nb/Sr) &(nb/Sr) \\
  \hline \\[0.05cm]
  \endfirsthead
  \hline
  \hline
$Q^2$ &$\epsilon$ &$W$ &$\cos{\cm{\theta}}$ &$\cm{\phi}$ &$d\sigma^{\gamma^{\ast}}$ &$\sigma_{stat}$ &$\sigma_{norm}$ &$\sigma_{sys}$ \\
(GeV$^2$) &none &(GeV) &none &(rad) &(nb/Sr) &(nb/Sr) &(nb/Sr) &(nb/Sr) \\
  \hline \\[0.05cm]
  \endhead
7.924 &0.226 &1.112 &-0.83 &-2.618 &3.3484 &3.3484 &0.19308 &0.00010623 \\[0.05cm]
7.924 &0.226 &1.112 &-0.83 &-1.571 &5.4915 &3.8833 &0.31666 &0.0015979 \\[0.05cm]
7.924 &0.226 &1.112 &-0.83 &0.5236 &3.6922 &3.6923 &0.2129 &0.0011096 \\[0.05cm]
7.924 &0.226 &1.112 &-0.5 &-2.618 &10.699 &6.1777 &0.61693 &0.00074719 \\[0.05cm]
7.924 &0.226 &1.112 &-0.5 &-1.571 &3.2228 &3.2229 &0.18584 &4.5554e-05 \\[0.05cm]
7.924 &0.226 &1.112 &-0.5 &0.5236 &13.981 &8.0726 &0.80618 &0.0011314 \\[0.05cm]
7.924 &0.226 &1.112 &-0.5 &2.618 &11.165 &6.4639 &0.64379 &0.00056412 \\[0.05cm]
7.924 &0.226 &1.112 &-0.17 &-2.618 &3.9838 &3.9839 &0.22972 &0.00011688 \\
\end{longtable}
 \end{center}
\end{widetext}

\begin{acknowledgments}
We would like to acknowledge the support of staff and management at
Jefferson Lab
\par
This work is supported in part by research grants from the
U.S. Department of Energy (including grant DE-AC02-06CH11357), the U.S.
National Science Foundation and the South African National Research Foundation.
\par
The Southeastern Universities Research Association operates the Thomas
Jefferson National Accelerator Facility under the U.S. Department of Energy
contract DEAC05-84ER40150.
\end{acknowledgments}

\bibliography{baryonprc}

\end{document}